\documentclass{aa}
\usepackage{multirow}
\usepackage{newtxtext,newtxmath}
\usepackage{booktabs}   

\usepackage[T1]{fontenc}

\DeclareRobustCommand{\VAN}[3]{#2}
\let\VANthebibliography\thebibliography
\def\thebibliography{\DeclareRobustCommand{\VAN}[3]{##3}\VANthebibliography}

\usepackage{graphicx}	
\usepackage{amsmath}	
\usepackage{subcaption}
\usepackage{url}
\usepackage{comment}
\usepackage{hyperref}

\newcommand{\cmfast}{{\tt 21cmFAST}}

\newcommand{\delsq}{\Delta^2_{\rm 21}}
\newcommand{\emuiii}{{\tt 21cmEMUv3}}

\newcommand{\Msun}{M_\odot}

\newcommand{\Muv}{M_{\rm 1500}}
\newcommand{\hone}{\mathrm{H}\textsc{i}}
\newcommand{\avenf}{\overline{x}_{\hone}}
\newcommand{\aveTs}{\overline{T}_{\rm S}}

\newcommand{\aveTb}{\overline{T}_{\rm b}}
\newcommand{\taue}{\tau_{\rm e}}

\begin{document}
   \title{\emuiii: a hybrid diffusion-LSTM emulator of \cmfast\ summary observables}

   \author{D. Breitman
          \inst{1,2,3}\thanks{E-mail: daniela.breitman@sns.it}
          \and
          A. Mesinger \inst{3,4}
          \and S. G. Murray \inst{3, 5}
          \and I. Nikolić \inst{6, 7, 3}
          \and R. Trotta \inst{8, 9, 10}
          }

   \institute{Research Center for the Early Universe, Graduate School of Science, The University of Tokyo, 7-3-1 Hongo, Bunkyo, Tokyo 113-0033, Japan \and
   Department of Physics, Graduate School of Science, The University of Tokyo, 7-3-1 Hongo, Bunkyo, Tokyo 133-0033, Japan
    \and
    Scuola Normale Superiore (SNS), Piazza dei Cavalieri 7, Pisa, PI, 56125, Italy
    \and
        Department of Physics and Astronomy {\it ``Ettore Majorana''}, University of Catania, Via Santa Sofia 64, 95123  Catania, Italy
             \and
             Physics Department, Stellenbosch University, 42 Merriman Ave, Stellenbosch, South Africa, 7600
             \and
             Cosmic Dawn Center (DAWN)
             \and
             Niels Bohr Institute, University of Copenhagen, Jagtvej 128, 2200 Copenhagen N, Denmark
             \and
             SISSA, Via Bonomea 265, 34136 Trieste, Italy \& INFN Sezione di Trieste
             \and 
            Centro Nazionale di Ricerca in High Performance Computing, Big Data e Quantum Computing, Via Magnanelli 2, Bologna, Italy
            \and 
            Physics Department, Blackett Lab, Imperial College London, Prince Consort Road, London SW7 2AZ, UK}
   \date{Accepted XXX. Received YYY; in original form ZZZ}

\abstract{

We are witnessing a surge in observations of the cosmic dawn (CD) and epoch of reionisation (EoR), driving an increasing demand for fast and robust theoretical interpretation frameworks. In response, machine learning (ML), and emulation in particular, has emerged as a powerful approach to accelerate and enhance inference pipelines. In this work, we present 21cmEMUv3, an emulator trained on 21cmFASTv3 simulations that model both atomically and molecularly cooling galaxies. 21cmEMUv3 is conditioned on $\sigma_8$ and ten astrophysical parameters to produce seven summary observables: (i) the cylindrical 21cm power spectrum (PS), emulated for the first time at such high resolution and accuracy across a wide redshift range of $z \sim$ 6--30; (ii) the spherically-averaged 21cm PS; (iii) the mean neutral fraction of the intergalactic medium (IGM); (iv) the mean 21cm spin temperature; (v) the global 21cm signal; (vi) the ultraviolet (UV) luminosity functions (LFs); and (vii) the Thomson scattering optical depth. Notably, the cylindrical 21cm PS is emulated via score-based diffusion, while the remaining six summaries are emulated via long-short term memory (LSTM) networks, all achieving sub-percent median accuracy. We use the emulator to reinterpret current 21cm PS upper limits from HERA, for the first time using state-of-the-art hydrodynamical simulations to inform priors on star formation inside molecularly cooling galaxies. We find that our inferred soft-band X-ray luminosity per unit star formation rate is consistent with extrapolations of high-mass X-ray binaries to the low-metallicity regimes expected in the first galaxies, excluding values below $10^{39.2}$ erg s$^{-1}M^{-1}_\odot \rm{yr}$ at $95\%$ confidence. Finally, we produce forecasts for the detection of the cosmic 21cm PS with the Square Kilometre Array for different array configurations. The 21cmEMU package is publicly available.
}

\titlerunning{\emuiii}
\maketitle


\section{Introduction}
\label{sec:intro}

One of the greatest challenges in modern cosmology is understanding the cosmic dawn (CD) of the first luminous sources and the subsequent reionisation of the intergalactic medium (IGM). Key open questions include when and how the first stars and galaxies formed, which sources drove reionisation, and how ionised bubbles grew and percolated through the IGM to complete the reionisation process. Over the past decade, there has been a surge in observational constraints to answer these questions, from a diverse suite of probes each sensitive to different aspects of this complex process. These include the Lyman-$\alpha$ forest (e.g., \citealt{Fan06,Becker07, Becker15,Bosman18,DOdorico23}), damping wings in quasar spectra (e.g., \citealt{Bolton11,Mortlock11,Banados18, Wang20, Yang20}), Lyman-$\alpha$ emission from galaxies (e.g., \citealt{Ouchi10,Clement12, Konno14, Drake17, Hoag19, Shibuya19}), large-scale polarization of the cosmic microwave background (CMB; e.g., \citealt{Planck18, DeBelsunce21, Heinrich21}, secondary kinetic Sunaev-Zeldovich (kSZ) CMB anisotropies (e.g., \citealt{Das14,George15,Reichardt21}), and finally upper limits on the cosmic 21-cm power spectrum (PS; e.g., \citealt{Mertens25,Nunhokee25,HERA25}).

 These observations are highly complementary and combining them in a standard Bayesian inference framework results in the strongest constraints on astrophysical and cosmological parameters (e.g., \citealt{Greig17, Qin20,HERA22,Breitman24, Nunhokee25, Mertens25}).   This is challenging however, as interpreting such diverse data requires large-scale cosmological simulations that self-consistently track galaxy evolution and how galactic radiation fields impact the IGM. Even the most efficient semi-numerical simulators (e.g., \citealt{Mesinger07, Visbal12,Ghara15,Maity22,Davies25, Santos10, Hutter21}) typically require $\gtrsim$ 1 core hour per forward model. This means that exploring the high-dimensional parameter space of astrophysical uncertainties can be very computationally expensive, even using semi-numerical simulations.  These computational challenges are further exacerbated by the fast pace of new EoR data releases, with each new dataset typically requiring a new inference.

As a result, there has been increasing interest in methods that accelerate Bayesian inference for EoR and CD data.  These typically "amortize" the computational cost: building an initial database to train a neural network (NN) to perform either emulation (e.g., \citealt{Kern17, Shimabukuro17}) or simulation based (i.e. implicit likelihood) inference (SBI; see e.g. \citealt{Cranmer19} for a review).  Once these NNs are trained, subsequent inferences can be performed at very low computational cost.


Amortized SBI involves stochastically sampling all relevant sources of uncertainty in order to build a training set of mock observation-parameter pairs that include realistic error realizations.  Then a NN is trained to rapidly evaluate the posterior (e.g. \citealt{Papamakarios16}), likelihood (e.g. \citealt{Papamakarios18}), or likelihood to evidence ratio (e.g. \citealt{Hermans20}), for the given observational data. 
As such, a key advantage of SBI is that it bypasses the need to explicitly define a likelihood,
which can be useful when the likelihood is analytically intractable.  
However, current EoR / CD data has relatively low signal-to-noise, and commonly used observational summaries tend to have nearly-Gaussian likelihoods (e.g. \citealt{Prelogovic23}).

Emulation, on the other hand, involves learning the mapping between model parameters and simulation outputs. Once trained, an emulator can be conveniently reused within an inference framework to interpret multiple datasets without retraining or with minimal refinement, assuming its output can be further post-processed to include instrument-specific effects. 
This flexibility makes emulation particularly well suited for broad community use and for keeping pace with frequent observational data releases. 
While in principle an emulator can be embedded within a likelihood-free framework, the emulator itself does not learn the likelihood in the way that SBI does. Consequently, the main drawback of emulation is that it requires an explicit specification of the likelihood function during inference, which may become an issue with future high signal-to-noise ratio (S/N) datasets.

Emulators were introduced into the CD/EoR community nearly a decade ago (e.g., \citealt{Kern17, Shimabukuro17, Schmit18, Ghara20}). Early efforts  focused on learning the 21-cm power spectrum, as it is the summary statistic targeted by interferometric 21-cm experiments. As interest grew in measuring the global (volume-averaged) 21-cm signal, emulators for this observable likewise became increasingly common (e.g., \citealt{Cohen20, Bevins21, Bye22}). Initial implementations typically relied on relatively simple architectures, such as Gaussian process regression (GPR; e.g., \citealt{Kern17}) and multilayer perceptrons (MLPs; \citealt{Shimabukuro17, Cooper25}). More recent emulators, however, have adopted substantially more sophisticated models, including convolutional neural networks (CNNs; e.g., \citealt{Breitman24,Scelfo26}), Bayesian neural networks (BNNs; e.g., \citealt{Mahida25}), recurrent neural networks (RNNs; e.g., \citealt{Prelogovic22, DorigoJones24}), generative adversarial networks (GANs; e.g., \citealt{Diao25}), and denoising diffusion probabilistic models (DDPMs; \citealt{Zhao23,Mishra25}) and transformers (e.g., \citealt{Moriwaki25}), as well as scattering transform-based maximum entropy generative models (e.g., \citealt{Hothi26}).

Despite this progress, most existing emulators are designed to predict a single observable. Previous studies (e.g. \citealt{HERA22a, HERA23, HERA25, Nunhokee25}) have shown that Bayesian inference with upper limits on the 21-cm power spectrum is significantly strengthened by a synergistic approach where multiple complementary observables are combined. This result created a rising interest in emulators capable of jointly predicting multiple observables within a unified framework.
In \cite{Breitman24}, we presented {\tt 21cmEMUv1}, the first emulator of six EoR summary statistics from the public semi-numerical simulator \cmfast\ \footnote{\url{https://github.com/21cmfast/21cmFAST}} \citep{Mesinger07, Mesinger11}. 
In \citealt{Cang24}, we introduced {\tt 21cmEMUv2} trained on a different astrophysical model which includes a population of molecularly-cooling galaxies (MCGs) (expected to host mostly Pop III stars and their remnants) in addition to the population of atomically-cooling galaxies (ACGs, hosting primarily Pop II stars) already present in {\tt 21cmEMUv1}. In this work, we present \emuiii, a significant expansion to previous versions both in terms of the physical model as well as the network architecture:
(i) \emuiii\ is trained on a \cmfast\ model including both populations of ACGs and MCGs, where we allow each population to have different stellar properties;
(ii) \emuiii\ is the first high-resolution emulator of the cylindrical (2D) 21-cm PS (c.f. \citealt{Cooper25}) with 32 bins along the sky-plane wavemode axis and 64 bins along the line-of-sight wavemode axis trained across a wide range of redshifts ($z \sim 5 - 30$)\footnote{The first 21-cm 2D PS emulator presented in \citealt{Cooper25} takes as input three astrophysical parameters and outputs a 21-cm 2D PS cube across eight redshifts $z \sim 5 - 12$ and eight bins along each wavemode axis. In comparison, \emuiii\ takes ten astrophysical parameters plus $\sigma_8$ along with a redshift $z \in [5,30]$ as an additional input parameter and outputs a 2D PS with 32 bins along the sky-plane wavemode axis and 64 bins along the line-of-sight wavemode axis. A finer Fourier-space grid more closely matches the native resolution of radio interferometers, avoiding information loss from heavy binning and enabling more direct comparison with observations. };
(iii) \emuiii\ is the first score-based diffusion \citep{Hyvarinen05, Sohl15, Song20} emulator of the \textit{mean} 21-cm PS (rather than a single realisation of the 21-cm PS);
(iv) \emuiii\ emulates five other summaries besides the 2D 21-cm PS with long-short term memory (LSTM) layers (e.g., \citealt{Hochreiter97}), whose purpose is to better capture temporal structure in timeseries data (e.g. \citealt{Prelogovic22, DorigoJones24}).

We begin by introducing the physical model being emulated in Section \ref{sec:model21cmfast}. Next, we describe the machine learning architecture of the emulator in Section \ref{sec:arch}. In Section \ref{sec:emu}, we quantify the performance of {\tt 21cmEMUv3} on a test set. In Section \ref{sec:hera_h1c}, we apply {\tt 21cmEMUv3} in an inference problem, where we demonstrate that the prior choice can notably affect the resulting posterior and therefore the corresponding scientific conclusions. We conclude in Section \ref{sec:conclusion}.
\section{Simulated database}
\label{sec:model21cmfast}




In this section we discuss the training set used to develop our emulator.  We begin by summarizing the galaxy model, and then discuss the cosmological simulations and emulated summary statistics.

\subsection{Galaxy model}
\label{sec:model}

Our simulations use the code {\tt 21cmFASTv3.0.4} \citep{Mesinger07,Mesinger11, Murray20}, with a galaxy model that includes two distinct galaxy populations: ACGs and MCGs.  For more details on the galaxy model, users are encouraged to consult \citealt{Qin20} and \citealt{Munoz22}.  Below we summarize the most relevant points, focusing on the free parameters that will serve as inputs to our emulator.


We characterize the stellar-to-halo mass relations (SHMRs) of the faint galaxies that dominate the EoR photon budget (e.g. \citealt{Qin25}) via power laws (e.g., \citealt{Behroozi15, Mutch16, Sun16, Yue16}).  Specifically, the stellar mass $M_{\ast, \textsc{acg (mcg)}}$ of an ACG (MCG) 
galaxy hosted by a halo of mass $M_h$ can be written as:
\begin{equation}
M_{\ast, \textsc{acg (mcg)}}/M_h = \min\left [ 1,  f_{\ast, 10 (7)} 
    \left( \frac{M_h}{M_{10 (7)}}\right)^{\alpha_{\ast, \textsc{acg (mcg)}}} \right] \times \left( \frac{\Omega_b}{\Omega_m}\right),
\end{equation}
where $\Omega_b$ and $\Omega_m$ are the baryon and total matter energy densities, respectively.
We can write the stellar fraction in ACGs (MCGs) as $f_{\ast, \textsc{acg (mcg)}} \equiv f_{\ast, 10 (7)} \left( \frac{M_h}{M_{10 (7)}}\right)^{\alpha_{\ast, \textsc{acg (mcg)}}}  \in [0,1]$, with $f_{\ast, 10(7)}$ corresponding to the fraction of galactic gas in stars normalized to the amount in a halo of mass $M_{10(7)} \equiv 10^{10(7)} M_\odot$.
These SHMRs introduce a total of four free parameters: the normalisation and power-law index of each of the two relations. We keep $f_{\ast, 10(7)}$ and $\alpha_{\ast, \textsc{acg}}$ as three free parameters, and fix $\alpha_{\ast} \equiv \alpha_{\ast, \textsc{mcg}} = \alpha_{\ast, \textsc{acg}}$ 9 (e.g. \citealt{Xu16}).

We assume that stellar mass grows exponentially over time. We write the corresponding star formation rate (SFR) as a function of the Hubble time, $H^{-1}(z)$, (or analogously the dynamical time, which also scales with the Hubble time during matter domination):
\begin{equation}
\Dot{M}_{\ast, \textsc{acg (mcg)}} = \frac{M_{\ast, \textsc{acg (mcg)}}}{ t_\ast H^{-1}(z)},
\end{equation}
where the characteristic star formation timescale, $t_\ast \in [0,1]$, is another free parameter.

Star formation is suppressed in low mass halos due to inefficient gas cooling and/or  feedback (e.g., \citealt{Hui97, Springel03, Okamoto08, Sobacchi13, Xu16, Ocvirk20, Ma20}). 
We account for this suppression by including an exponential duty cycle to the mass function of halos (HMF) that host star-forming ACGs (MCGs):
\begin{equation} 
    \phi^{\textsc{acg (mcg)}} = \frac{{\rm d}n}{{\rm d} M_h} \times
    \begin{cases}
          \exp \left( - \frac{M_{\rm crit}^{\textsc{acg}}}{M_h}\right)\\
          \exp \left( - \frac{M_{\rm crit}^{\textsc{mcg}}}{M_h}\right) \exp \left( - \frac{M_h}{M_{\rm crit}^{\rm cool}}\right)
        \end{cases}.
\end{equation}
$M_{\rm crit}^{\textsc{acg (mcg)}}$ is the turnover mass below which star formation is suppressed, which we write as:
\begin{equation}
    M_{\rm crit}^{\textsc{acg (mcg)}} = \max \left[M_{\rm crit}^{\rm cool \: (diss)}, M_{\rm crit}^{\rm ion}, M_{\rm crit}^{\textsc{sn}} \right].
\end{equation}
Here the efficiency of star formation is limited by three physical processes: (i) inefficient cooling $M_{\rm crit}^{\rm cool}$ below the atomic cooling threshold for ACGs (e.g. \citealt{Mesinger07}) and Lyman-Werner feedback that photodissociates molecular hydrogen via the two-step Solomon process, $M_{\rm crit}^{\rm diss}$, for MCGs \citealt{Haiman00,Visbal14, Munoz22}); (ii) photoheating feedback, $M_{\rm crit}^{\rm ion}$, inside the ionized IGM (e.g. \citealt{Okamoto08,Sobacchi13}); (iii) supernova feedback, $M_{\rm crit}^{\textsc{sn}}$ (cites).
Moreover, for MCGs, in addition to the lower mass threshold, $M_{\rm crit}^{\textsc{mcg}}$, we also include a second exponential term with an upper mass threshold, $M_{\rm crit}^{\rm cool}$, to smoothly transition between MCGs and ACGs around the atomic cooling threshold $T_{\rm vir} \sim 10^4 {\rm K}$ (c.f. \citealt{Qin20}).

The population-averaged ionizing escape fraction, $f_{\rm esc, \textsc{acg (mcg)}}\in [0,1]$ is described by a power-law (e.g., \citealt{Paardekooper15,Kimm17, Lewis20}):
\begin{equation}
f_{\rm esc, \textsc{acg (mcg)}} = \min \left[1, f_{\rm esc, 10 (7)} \left( \frac{M_h}{M_{10 (7)}} \right)^{\alpha_{\rm esc}}\right],
\end{equation}
where we have a total of three free parameters for both relations: the normalizations $f_{\rm esc, 10 (7)}$, and the power-law index, $\alpha_{\rm esc}$.

The specific X-ray luminosity escaping the galaxies is also taken to be a power-law in energy (e.g., \citealt{Das17}), $L_{X,\textsc{acg (mcg)}} \propto E^{-\alpha_X}$. We normalise it via the soft-band (i.e. $< 2$ keV) X-ray luminosity per unit SFR for each galaxy population:
\begin{equation}
    L_{X<2 {\rm keV/}}^{\textsc{acg (mcg)}} / \Dot{M}_{\ast, \textsc{acg (mcg)}} = \int_{E_0}^{2\rm{keV}} {\rm dE} \ L_{\textsc{x}} / \Dot{M}_{\ast, \textsc{acg (mcg)}},
\end{equation}
where $E_0$ is the minimum energy of X-ray photons capable of escaping their host galaxy, for a total of three additional free parameters.

\begin{table*}[ht]
\caption{In \emuiii, we model ACGs and MCGs by varying a total of eleven parameters: seven ACG astrophysical parameters, three MCG astrophysical parameters, and cosmological parameter $\sigma_8$. The second column shows the truncated Gaussian distribution mean, standard deviation, and the upper and lower bounds for each parameter. The ACG, MCG, and $\sigma_8$ priors are motivated by \citep{Qin21, Park19}, \citep{Qin21b}, and \citep{Planck18}, respectively.}
\label{tab:params}
\begin{tabular}{lll}
Parameter            & Distribution and limits & Description \\ \hline
$\log_{10} f_{*,10}$       & $\mathcal{N}(-1.2,0.2)\in [-2, -0.5]$     & SHMR normalisation for ACGs          \\
$\alpha_{\ast}$   & $\mathcal{N}(0.5,0.15)\in[0,1]$   & power-law index of the SHMR       \\
$t_{\ast}$         & $\mathcal{N}(0.55,0.3)\in[0.01,1]$    & characteristic SF timescale         \\
$\log_{10} f_{\rm esc, 10}$     & $\mathcal{N}(-1.3,0.4)\in[-3,0]$   & normalisation of the $f_{\rm esc}-\rm{M}_h$ relation for ACGs         \\
$\alpha_{\rm esc}$     & $\mathcal{N}(0,0.5)\in[-1,1]$       & power-law index of the $f_{\rm esc}-\rm{M}_h$ relation          \\
$\log_{10} f_{\ast,7}$ &  $\mathcal{N}(-2.5,0.8)\in[-4,-1]$      & SHMR normalisation for MCGs   \\
$\log_{10} f_{\rm esc, 7}$ & $\mathcal{N}(-1.5,0.8)\in[-3,-1]$& normalisation of the $f_{\rm esc}-\rm{M}_h$ relation for MCGs \\
$\log_{10}L_{\rm X<2keV}^{\textsc{acg}}/{\rm SFR}$ & $\mathcal{N}(40.5,1)\in[38,43]$ $\frac{{\rm erg}\: {\rm s}^{-1}}{M_\odot \: {\rm yr}^{-1}}$ & soft-band X-ray luminosity per unit SFR for ACGs\\
$\log_{10}L_{\rm X<2keV}^{\textsc{mcg}}/{\rm SFR}$ & $\mathcal{N}(41.5,1)\in[39,44]$ $\frac{{\rm erg}\: {\rm s}^{-1}}{M_\odot \: {\rm yr}^{-1}}$ & soft-band X-ray luminosity per unit SFR for MCGs \\
$E_{\rm 0}$ & $\mathcal{N}(500,300)\in[100,1500]$ keV & minimum energy of X-ray photons that can escape their host galaxy\\
$\sigma_8$ & $\mathcal{N}(0.8118,0.01)\in[0.75,0.85]$& matter power spectrum normalisation\\
\end{tabular}

\end{table*}

In \emuiii, we model ACGs and MCGs by varying a total of eleven parameters: seven ACG astrophysical parameters, three MCG astrophysical parameters, and one cosmological parameter $\sigma_8$. In Table \ref{tab:params}, we list the \emuiii\ input parameters and their limiting values in the database used to train the emulator (c.f. Section \ref{sec:db}).

\subsection{Simulations and summary observables}
\label{sec:summaries}
For a set of cosmological and astrophysical parameters, \cmfast\ produces 3D lightcones of IGM properties. When performing inference, these lightcones are generally compressed into summary statistics that are compared with observations. As such, rather than emulating full 3D lightcones, we emulate only lower-dimensional summary statistics:
\begin{enumerate}[(i)]
    \item $\boldsymbol{\overline{x}_{\textsc{hi}}(z)}$ --- volume-averaged (global) neutral fraction of hydrogen as a function of redshift (aka the EoR history).
    \item $\boldsymbol{\overline{T}_{\textsc{s}}(z)}$ --- the mean neutral IGM spin temperature as a function of redshift \footnote{   
    The IGM spin temperature is only defined for neutral hydrogen that is outside of the cosmic HII regions that surround galaxies. As in \texttt{21cmEMUv1}, the volume average is performed over simulation cells that have $x_{\rm HI} \geq$ 95 \%.}.
    \item $\boldsymbol{\overline{\delta T}_{\rm b}(z)}$ --- global 21-cm brightness temperature (e.g., \citealt{Madau97, Furlanetto06, Pritchard12}): 
\begin{align}
\label{eq:brightness_temperature}
    \delta T_{\rm b}(\boldsymbol{x}, z) &=  \frac{T_{\rm S} - T_{\rm R}}{1+z}(1-e^{\tau_{21}})\\ 
\nonumber &\approx  27 \ x_{\rm HI} (1 + \delta_b) \left( \frac{\Omega_b h^2}{0.023}\right) \left( \frac{0.15}{\Omega_m h^2} \frac{1+z}{10}\right)^{1/2} \ \rm{mK} \\
\nonumber &\times \left( \frac{T_{\rm S} - T_{\rm R}}{T_{\rm S}}\right) \left[ \frac{\partial_r v_r} {(1+z)H(z)}\right],
\end{align}
where $\tau_{21}$ is the 21-cm optical depth of the intervening gas, $\delta_b \equiv \rho/\bar{\rho} - 1$ is the baryon overdensity, with $\rho$ being the baryon density, and $T_{\rm S}$ and $T_{\rm R}$ are the spin and background temperatures, respectively.  We assume throughout that the radio background is provided by the CMB, $T_{\rm R} = T_{\rm CMB}$ is the temperature of the CMB.  We note that \cmfast\ computes the brightness temperature at each cell location, ${\bf x}$, using the exact expression in the first line of the equation above; the second line is a Taylor expansion in the limit of $\tau_{21} \ll 1$ that provides physical intuition.
    \item $\boldsymbol{\Delta^2_{21} (k_\perp, k_\parallel,z)}$ --- cylindrically-averaged 21-cm power spectrum (PS) as a function of sky-plane mode $k_\perp$ and line-of-sight (LOS) mode $k_\parallel$: $\Delta^2_{21}(k_\perp, k_\parallel,z)\left[\rm{mK}^2\right] \equiv k^3/(2 \pi^2) \langle \tilde{T}_{b} \tilde{T}_{b}^\ast\rangle$, where $k = \sqrt{k^2_\perp + k^2_\parallel}$, and $\tilde{T}_{b}(\boldsymbol{k}, z)$ is the Fourier dual of the brightness temperature from eq. (\ref{eq:brightness_temperature}). Note that in this work we calculate all PS with {\tt tuesday}\footnote{\url{github.com/21cmfast/tuesday}} \citep{Breitman25} and use trilinear interpolation to angular grid-points when performing angular averaging, instead of simple standard binning. This approach avoids biases that may occur when the signal evolves over the width of the radial bin.
    \item $\boldsymbol{\Delta^2_{21} (k,z)}$ --- spherically-averaged 21-cm PS as a function of wavemode $k$ \footnote{We emulate the 1D PS separately rather than averaging over the cylindrical 21-cm PS to minimise the uncertainty and speed up the emulation. Averaging the 2D PS uses score-based diffusion and is therefore slower. Since the 2D PS is a much more complex quantity than the 1D PS, averaging it down to 1D yields a larger error than emulating the 1D PS directly. Users may employ either method depending on the desired application.}.
    \item $\boldsymbol{\phi({\rm M}_{1500}, z)}$ --- the non-ionizing UV luminosity function (UV LF), defined as the number density of galaxies per UV magnitude, $\Muv$, as a function of redshift.  The $\sim$1500 \AA\ rest frame luminosity is assumed to be proportional to the SFR:
$L_{\rm UV} = \dot{M}_\ast (M_h,z) / \mathcal{K}_{\rm UV},$ where  $\mathcal{K}_{\rm UV} = 1.15 \cdot 10^{-28} \rm{M}_\odot \rm{ yr}^{-1} \ \rm{Hz} \ \rm{s}\ {erg}^{-1}$ assumes a Salpeter initial mass function (e.g., \citealt{Madau14, Sun16}).
The UV luminosity is related to the AB magnitude using \citep{Oke83}: $\log\left( \frac{L_{\rm UV}}{\rm{erg} \ {s}^{-1} \ {Hz}^{-1}}\right) = 0.4 \times (51.63 - M_{\rm UV}).$
    \item $\boldsymbol{\tau_e}$ --- the CMB optical depth:
$\tau_e = \sigma_T \int_0^{z_{\rm{LSS}}} dz \ \vline \frac{c dt}{dz} \vline \ n_e $,
where $\sigma_T$ is the Thompson scattering cross section, $z_{\rm LSS} \sim 1100$ the redshift of lass scattering, and $n_e$ is the electron number density calculated assuming hydrogen and helium are singly ionized at a fraction $(1-\avenf)$ and that helium is doubly ionized at $z < 3$.
\end{enumerate}



\subsection{Database}
\label{sec:db}
We create a database of $\sim$47k simulations total: $\sim$44k with standard sampling and an additional $\sim$3k with higher SHMRs, as described below. Each simulation has unique initial conditions, consisting of a 400 cMpc lightcone spanning redshift $z \sim 5-35$ at a resolution of 1.5 cMpc/cell. Rather than sampling parameters uniformly, we populate the database by sampling from truncated Gaussian distributions whose mean, standard deviation, and bounds are shown in Table \ref{tab:params}. 

In Table \ref{tab:params}, the means and standard deviations of the prior distributions of each ACG parameter (namely, $f_{\ast,10}$, $\alpha_\ast$, $t_\ast$, $f_{\rm esc,10}$ and $\alpha_{\rm esc}$) are motivated by results from \citealt{Qin21}, who performed a joint Bayesian inference combining high-redshift UV luminosity functions \citep{Bouwens15, Bouwens16, Oesch18}, the neutral hydrogen fraction from quasar dark pixel fractions \citep{McGreer15}, the CMB Thomson scattering optical depth \citep{Planck18, Qin20}, and Lyman-$\alpha$ forest opacity fluctuations \citep{Bosman18}. The prior for the X-ray parameters ($L_{\rm X<2keV}^{\textsc{acg}}/{\rm SFR}$ and $E_0$) are based on \citet{Park19}. The prior for the three MCG parameters $\log_{10} f_{\ast,7}$, $\log_{10} f_{\rm esc, 7}$, and $\log_{10}L_{\rm X<2keV}^{\textsc{mcg}}/{\rm SFR}$  are based on \citet{Qin21b}, where the X-ray production parameter, $L_{\rm X<2keV}^{\textsc{mcg}}/{\rm SFR}$ is adjusted to allow for higher X-ray luminosities at earlier cosmic times as implied by \citet{HERA22}. The $\sigma_8$ prior is based on the posterior analysis in \citet{Planck18}. Centring the sampling distributions around these values ensures that the database is densely sampled in  regions of parameter space most consistent with current observations, rather than sampling uniformly across the full prior volume.

We find below however that the resulting prior is insufficiently broad to cover the full SHMR range required for the application in Section \ref{sec:hera_h1c}. We therefore supplement the database with an additional $\sim$3k simulations with higher SHMRs, in which $\log_{10} f_{*,10}$ and $\alpha_\ast$ are drawn from $\mathcal{N}(0,0.3)\in[-1,1]$ and $\mathcal{N}(0.9,0.08)\in[0.5,1.2]$, respectively, with all remaining parameters retaining the distributions in Table \ref{tab:params}. 



\section{21cmEMUv3 architecture and training}
\label{sec:arch}
\begin{figure*}
    \centering
    \includegraphics[width=0.75\linewidth]{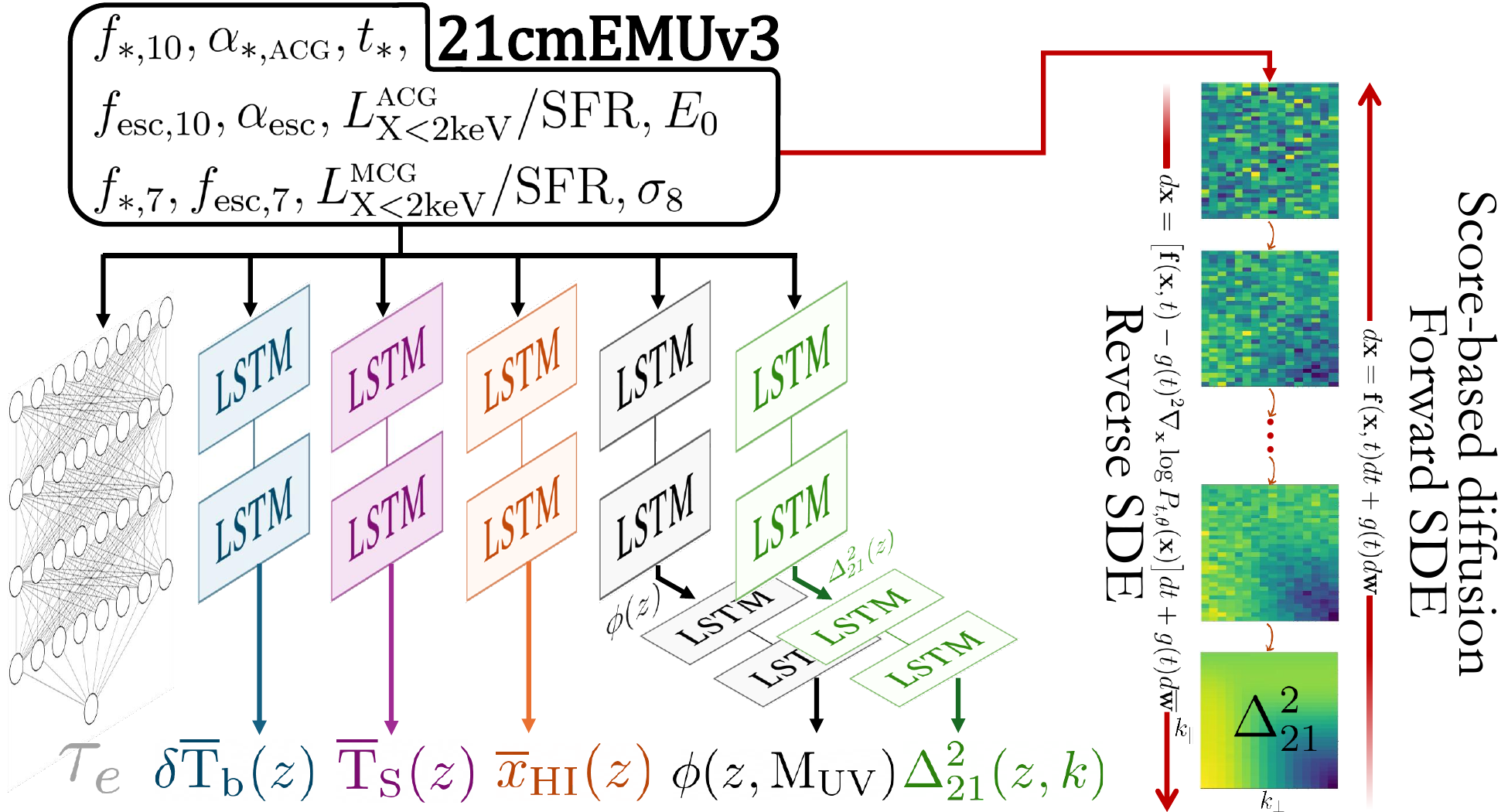}
    \caption{Diagram of the architecture of \texttt{21cmEMUv3}. The input consists of a total of eleven parameters: $\sigma_8$ and ten astrophysical parameters, three of which affect pop III galaxies. \texttt{21cmEMUv3} consists of two networks: timeseries LSTM network on the left and the score-base diffusion network for the 2D 21-cm PS on the right. The network on the left emulates the globally-averaged 21-cm signal, spin temperature, and neutral fraction all as a function of redshift $z$. The UV LFs as a function of magnitude at seven redshifts and spherically-averaged PS as a function of redshift and wavemode are emulated with 2D LSTM networks since they are two-dimensional quantities. This network also includes a branch with a feed-forward neural network that emulates the Thomson scattering optical depth. The eleven parameters along with the desired redshift are also passed to the score-based diffusion network that produces one 2D 21-cm PS in $\sim2$s on a single GPU.}
    \label{fig:diagram_full}
\end{figure*}

Figure \ref{fig:diagram_full} shows the architecture of {\tt 21cmEMUv3}. It is composed of two independent parts: the generative score-based diffusion model for the cylindrical 21-cm PS (right branch in red in Figure \ref{fig:diagram_full}), and the LSTM plus multilayer perceptron (MLP) architecture (left branch in black in Figure \ref{fig:diagram_full}) that emulates the remaining six summary observables. In this section, we describe the architecture of each of these neural networks.

\subsection{Generative score-based diffusion emulator of the cylindrical 21-cm PS}
\label{sec:sbm}

State-of-the-art CD/EoR simulations of the 21 cm signal typically span volumes that are orders of magnitude smaller than those probed by observations, primarily because such simulations become computationally impractical as their size increases. As a result, the number and overlap of Fourier modes common to both simulation and observation is limited, particularly in the cylindrical (2D) k-space that is natural for 21 cm interferometry. This mismatch introduces a key challenge for interpreting upcoming 21-cm observations at larger scales: sample variance, the deviation of the simulated realisation from the true population mean due to the finite simulation volume. Following \citealt{Breitman25}, we treat sample variance as mildly non-Gaussian noise \citep{Mondal15, Shaw19} added onto the target mean PS, and use score-based diffusion \citep{Song20, Ho20} to marginalise over it. In the rightmost branch of Figure \ref{fig:diagram_full}, we illustrate the diffusion process as applied to a cylindrical 2D PS:
\begin{itemize}
    \item Bottom to top: the forward diffusion process is a continuous noise-adding stochastic process where we corrupt a 2D PS realisation from the training sample with Gaussian noise with increasing variance until it is transformed into a sample from a known Gaussian prior distribution. This process can be described with the following stochastic differential equation (SDE): 
\begin{equation}
    d\mathbf{x} = \mathbf{f}(\mathbf{x},t) dt + g(t) d\mathbf{w},
\end{equation}
where $\mathbf{x} = \big(x(0), \hdots, x(T)\big)$ is the diffused 2D PS realisation at a given time in the diffusion process, $t \in [0,T]$, with $x(0)$ denoting a sample from the data distribution and $x(T)$ a sample from the Gaussian prior. $\mathbf{w}$ is a Wiener process (aka Brownian motion). The choice of drift function $\mathbf{f}(\mathbf{x},t)$ and diffusion coefficient $g(t)$ are essentially model hyperparameters. As in \citealt{Breitman25}, we choose the drift function and diffusion coefficient that correspond to the commonly-used variance preserving (VP) SDE, the continuous time-limit of the DDPM in \citealt{Ho20}. In this work, however, we also test two other common SDEs: the subVP SDE first introduced in \citealt{Song20} and the variance exploding (VE) SDE. We find that the emulator performs best with the VP SDE.
\item Top to bottom: to emulate the cylindrical PS, we need to reverse the forward process while conditioning it on the desired input parameters $\theta$. This forward process can be reversed with another SDE \citep{Anderson82}:
\begin{equation}
    d\mathbf{x} = \big[\mathbf{f}(\mathbf{x},t) - g(t)^2 \nabla_{\mathbf{x}} \log P_{t, \theta} (\mathbf{x}) \big] dt + g(t) d\mathbf{\overline{w}},
\end{equation}
where the only unknown is the \textit{score function} $\nabla_{\mathbf{x}} \log P_{t, \theta} (\mathbf{x})$ of the probability density function (PDF) $P$ of the data $x$ explicitly conditioned on the eleven input parameters and the redshift all denoted by $\theta$, in addition to the continuous time index $t$. 
\end{itemize}
The neural network is therefore trained to learn the score function in order to solve the reverse SDE, and conditioning it on the eleven input parameters $\theta$ allows us to generate new 2D PS samples for any parameter combination.

For the NN architecture, we choose a standard U-Net \citep{Ronneberger15} autoencoder\footnote{The model architecture is based on the {\tt PyTorch} implementation available here: \url{https://github.com/lucidrains/denoising-diffusion-pytorch} that is in turn based on the original implementation from \citealt{Ho20} here: \url{https://github.com/hojonathanho/diffusion}} following \citealt{Song20, Ho20}, as U-Nets are efficient in recognising local information in images over a range of scales. The model is implemented with {\tt PyTorch} \citep{pytorch}. The network is trained on 44k 2D PS at 32 redshifts logarithmically-spaced between 5.5 and 29 and are split into training (90\%) and validation (10\%) for $\sim$500 epochs using the \texttt{AdamW} \citep{Loshchilov19} optimiser with an initial learning rate of $10^{-4}$, a cosine annealing schedule with $\eta_{\rm min} = 10^{-6}$, and exponential moving average (EMA) checkpointing. 
Since we plan to integrate this emulator into an inference pipeline, we solve the reverse SDE via the fast probability-flow ODE \citep{Song20} method rather than Euler-Maruyama, emulating a mean 2D PS averaged over 200 samples in $\sim$2 s on a GPU. Our entire code is publicly available\footnote{\url{https://https://github.com/21cmfast/21cmEMU/releases/tag/v3.0}}. 

Each training sample is a single 2D PS realisation that can be interpreted as a mean PS with sample variance added as stochastic noise. Since the "noise" (i.e. initial conditions) realisation is unique to each simulation, learning any specific noise pattern would be penalised by the loss across training examples at similar $\theta$. The model therefore implicitly learns to associate parameters with the underlying mean PS while treating sample variance as noise. As a result, individual diffusion samples already approximate the mean PS. Averaging over multiple diffusion samples at fixed $\theta$ further suppresses the residual generated variance, yielding predictions that minimally suffer from sample variance in contrast to the individual training realisations, as we shall see in Section \ref{sec:ps}.

\subsection{LSTM emulator for timeseries data}
\label{sec:lstm}

A typical feed-forward neural network processes information in a single direction. Long short-term memory (LSTM) networks \citep{Hochreiter97} are a type of recurrent neural network (RNN; e.g. \citealt{Rumelhart86, Werbos90}), a broader class of architectures that, in contrast to feed-forward NNs, incorporate internal feedback loops to efficiently learn patterns in sequential data. Unlike feed-forward networks, RNNs maintain an internal state or "memory" that allows information from previous inputs to influence subsequent predictions. As such, RNNs are particularly well-suited for sequential data where context is essential, such as natural language processing and timeseries data (see e.g. \citealt{Schmidt19, Mienye24} for recent reviews). 

We adopt LSTMs following \citealt{DorigoJones24}, who found them well-suited for emulating the global 21-cm signal, and building on the earlier finding of \citealt{Prelogovic22} that RNNs perform well on 21-cm lightcone regression. The 2D outputs (1D PS and UV LFs) use an 2D LSTM-based head designed to capture correlations across both redshift and wavenumber/magnitude axes, which we find outperforms both a standard 1D LSTM head and a CNN head for these outputs. The scalar Thomson optical depth is emulated with a simple feed-forward network with four layers of 256 nodes each. All LSTM branches use two layers, as additional layers do not improve performance, consistent with \citealt{DorigoJones24}. 

Training this multi-output network presents a multi-task learning challenge, as the 1D PS is significantly more complex than the other summaries and benefits from being trained jointly rather than in isolation.  Similarly to the 2D PS, we also find that the LSTM naturally marginalises over sample variance and tends to emulate the mean PS, as expected with standard mean-squared error (MSE) loss. We explored several strategies including gradient surgery methods such as CAGrad \citep{Liu21} and various loss weighting schemes, finding only marginal improvements over a simpler baseline. Moreover, we find that weight decay is catastrophic for LSTM performance and do not use it. We therefore adopt a straightforward training procedure: the \texttt{Adam} optimiser \citep{Kingma14} with a \texttt{ReduceLROnPlateau} scheduler, decaying from $10^{-3}$ to $10^{-7}$, with EMA checkpointing. Training converges by $\sim$400 epochs on $\sim$45k training samples.

\section{21cmEMUv3 performance}
\label{sec:emu}
We assess the performance of \emuiii\ by evaluating it on a test set and calculating the error using the following metrics: 
\begin{enumerate}[(i)]
    \item absolute difference (Abs Diff): 
    \begin{equation}
        \text{Abs Diff} \equiv | y_{\rm true} - y_{\rm pred} |
    \end{equation}
    \item fractional error (FE):
    \begin{equation}
        {\rm FE} (\%) = \frac{\text{Abs Diff}}{\max (|y_{\rm true}|, y_{\rm floor})} \times 100,
    \end{equation}
    where the floor is set to $y_{\rm floor} = 1.0$ for the 21-cm PS in log space ($y_{\rm floor} = 10^{-2}$ mK$^2$ for the 21-cm PS in linear space), and neutral fraction, $y_{\rm floor} = 5$ mK for the global 21-cm signal. No floor is applied to other summaries.
\end{enumerate}

\subsection{Cylindrical 21-cm PS}
\label{sec:ps}
Training an emulator of the cylindrical PS is significantly more challenging than for the 1D PS due to the increased dimensionality. We find that emulator performance varies by $\sim$10--20\% when the training set size changes by a similar amount. To maximise accuracy, we use the $\sim$90\% of the database for training and $\sim$10\% for validation, reserving only 90 parameter combinations for the test set.

\begin{figure*}
    \centering
    \includegraphics[width=0.75\linewidth]{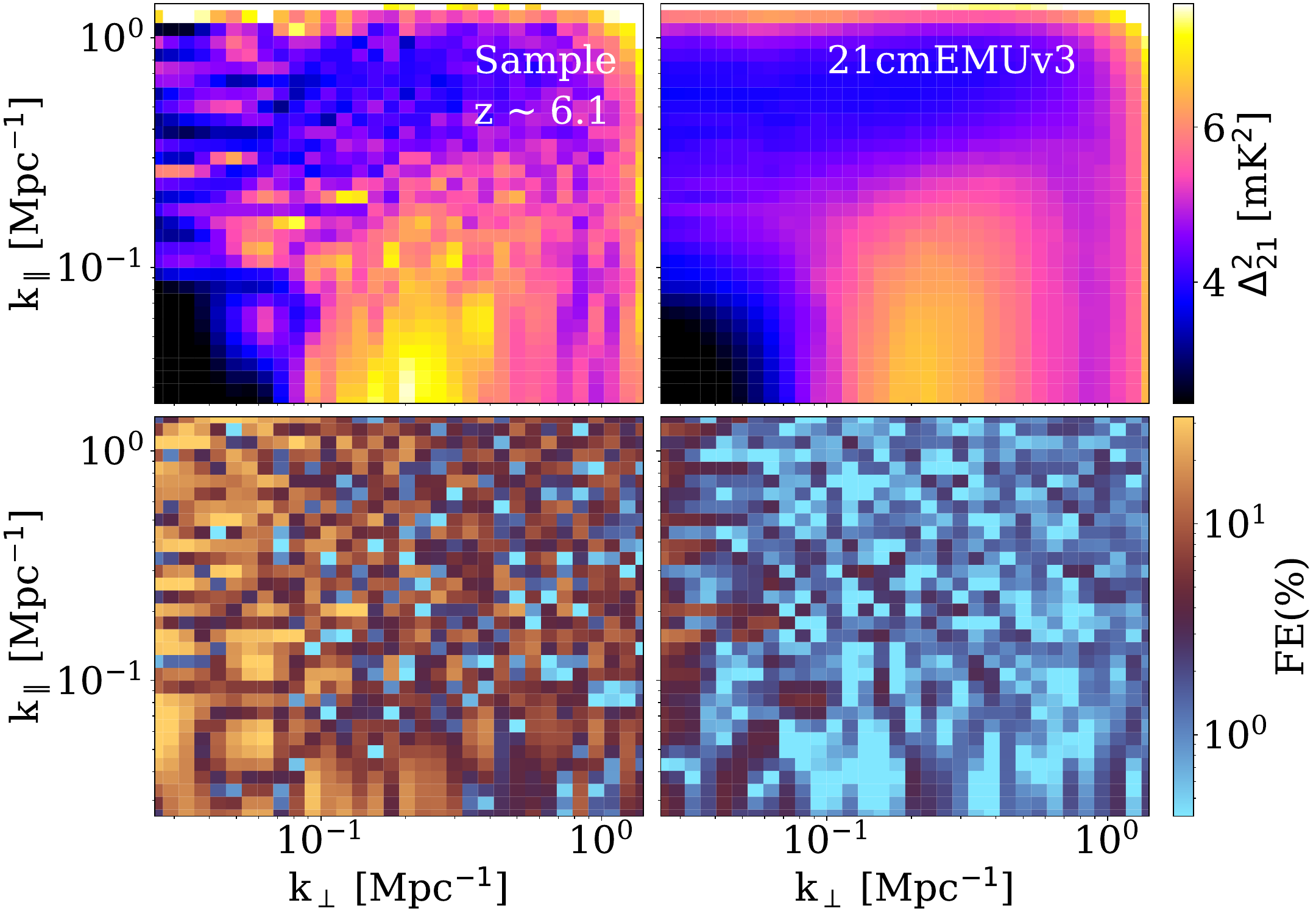}
    \caption{Top row: a cylindrical 21-cm PS realisation from the test set (left) and the corresponding \emuiii\ output (right). Bottom row: fractional error comparing each PS in the top row against the mean PS estimated by averaging 76 IC realisations at fixed astrophysical and cosmological parameters. The emulator produces a 2D PS much closer to the mean than the single realisation.}
    \label{fig:2d_ps_example}
\end{figure*}

As mentioned in Sections \ref{sec:sbm} and \ref{sec:lstm}, we expect {\tt 21cmEMUv3} to naturally marginalise over sample variance in both the 1D and 2D PS, since each training sample has a different Gaussian IC realisation and learning the mean minimises the loss. We check this by simulating an average of $\sim$60 PS realisations (varying the IC seeds) for each of the 90 parameter combinations in the test set and averaging over the realisations to obtain mean PS estimates. Comparing against the IC-averaged PS, we find that the emulator error is smaller than the sample variance error of individual realizations (by $\sim$7\% and $\sim$5\% for the 2D and 1D PS respectively). Figure \ref{fig:2d_ps_example} illustrates this for a single random sample of the 2D PS at redshift $z\sim6$ (top left), showing that the emulated PS (top right) deviates from the mean by less than $\sim$1\% across most scales (bottom right), while a single realisation may deviate by over $\sim$10\% (bottom left). In Appendix \ref{sec:emu_errs}, Figure~\ref{fig:FE_corner} shows the linear 2D PS fractional error as a function of each of the eleven input parameters across the 90-parameter test set, confirming that the $\lesssim$1\% accuracy is maintained uniformly across the full parameter range and does not degrade toward the prior edges.

\subsection{Other summaries}
\label{sec:timeseries}

\begin{figure*}
\begin{subfigure}{\linewidth}
    \includegraphics[width = 0.49\columnwidth]{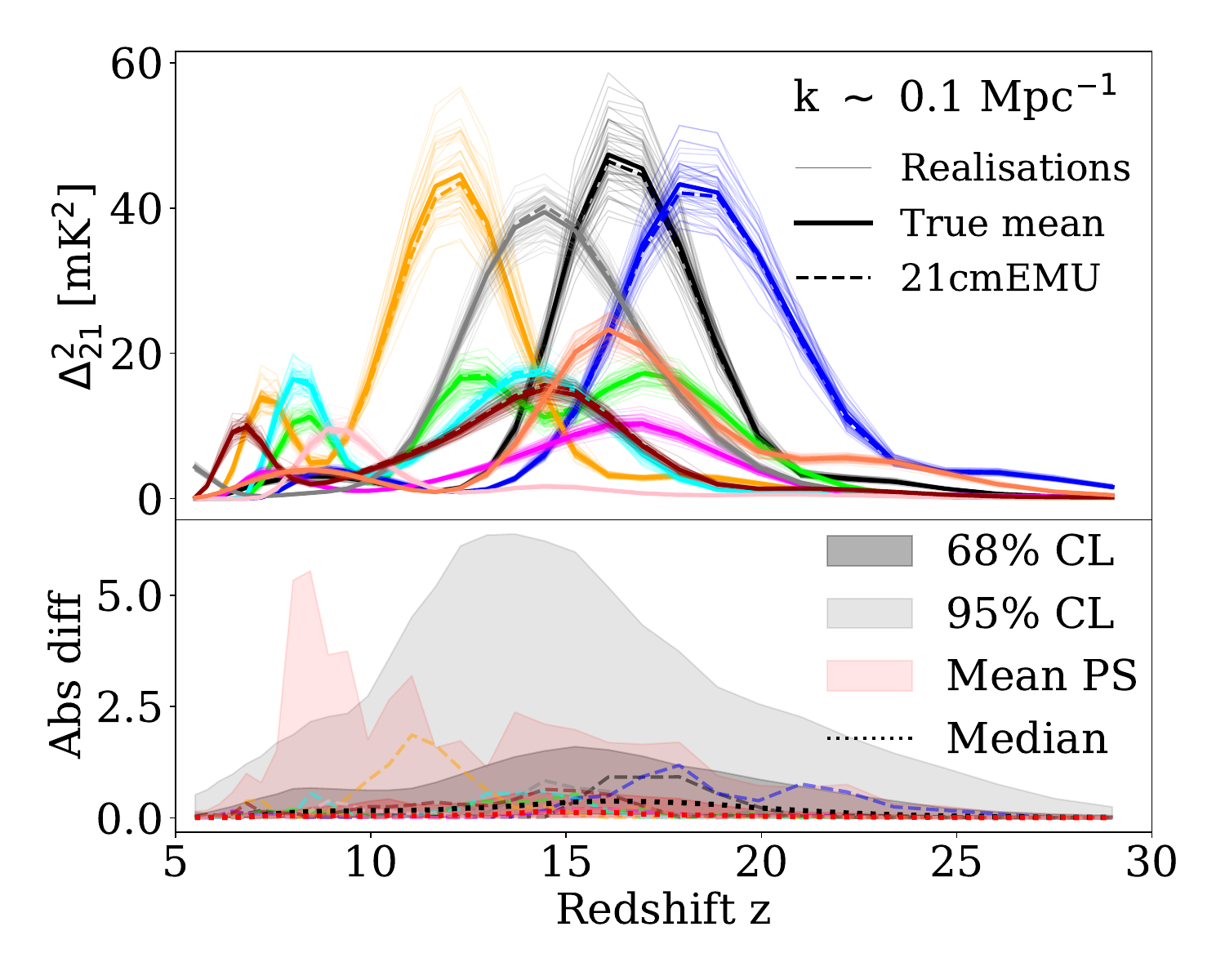} \hfill
    \includegraphics[width = 0.49\columnwidth]{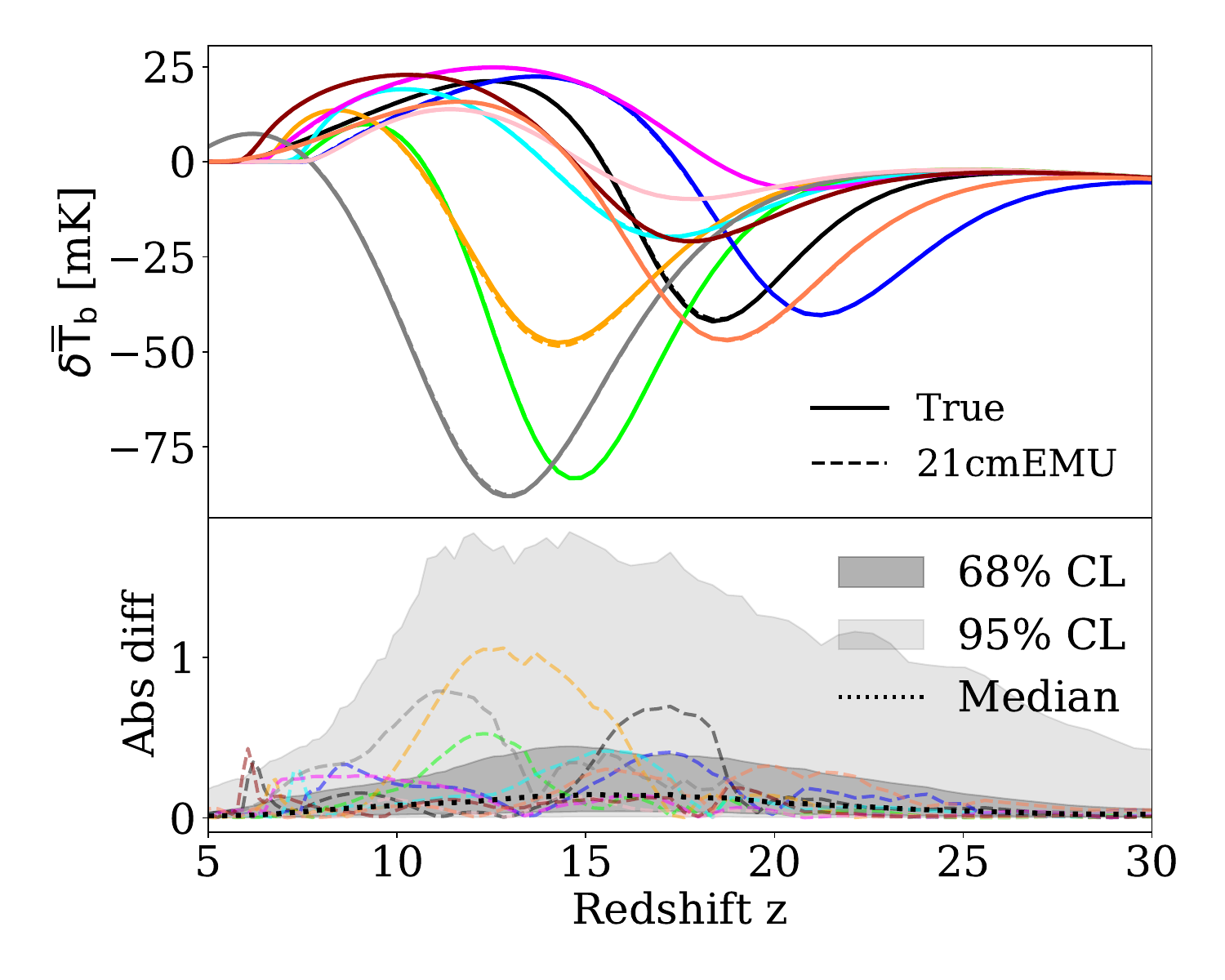} \\

    \includegraphics[width = 0.49\columnwidth]{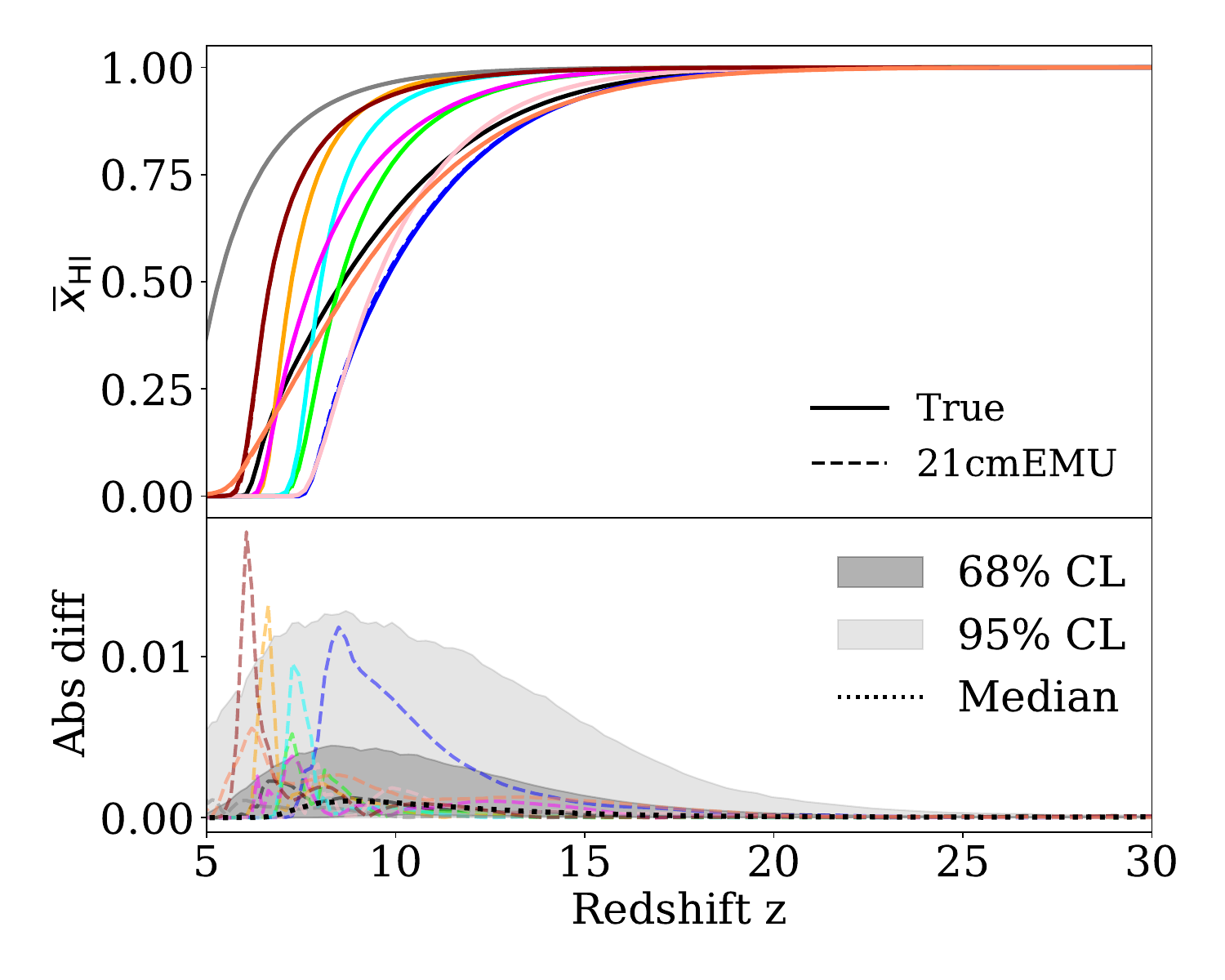}\hfill
    \includegraphics[width = 0.49\columnwidth]{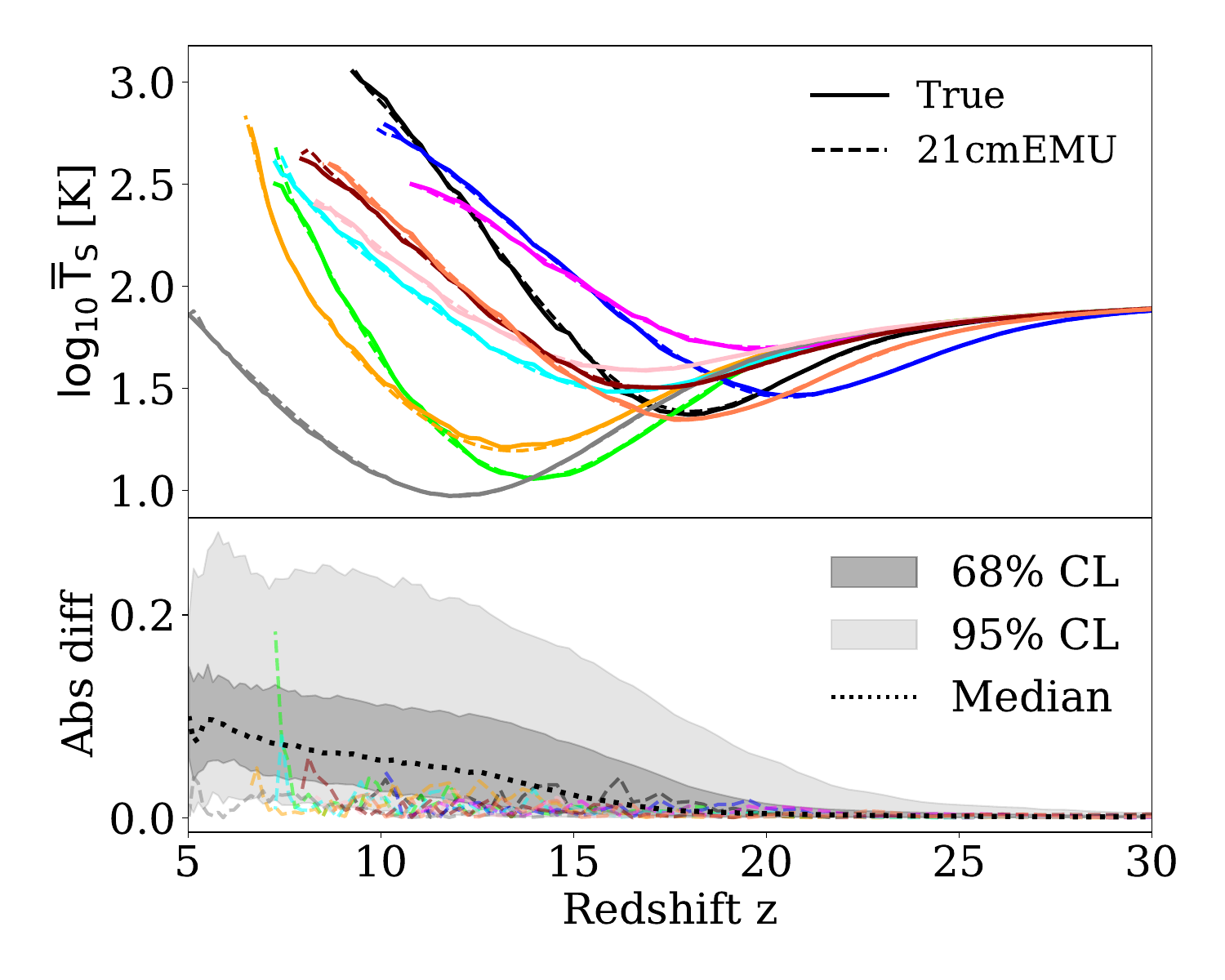} \\
    \includegraphics[width = 0.49\columnwidth]{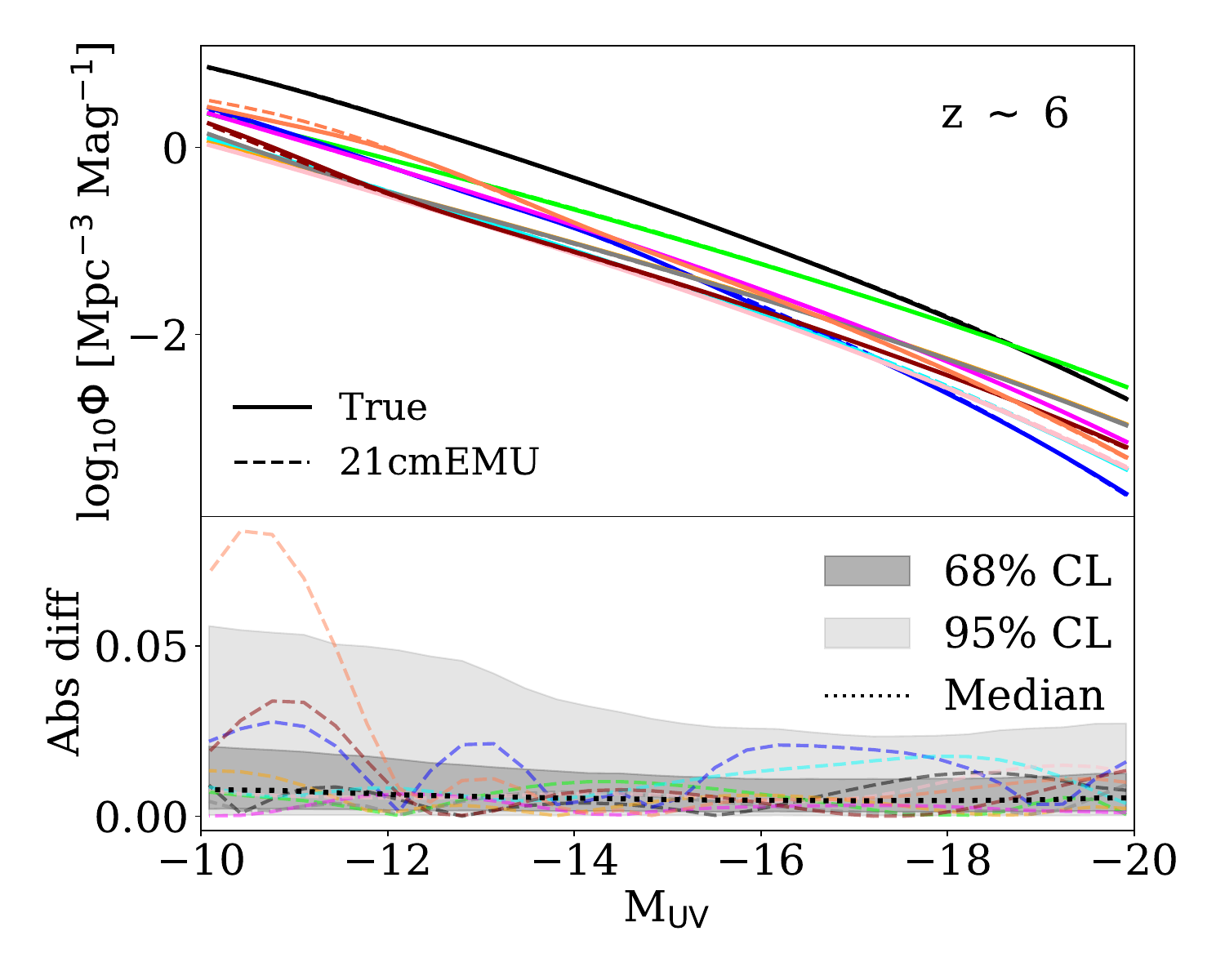} \hfill
    \includegraphics[width = 0.49\columnwidth]{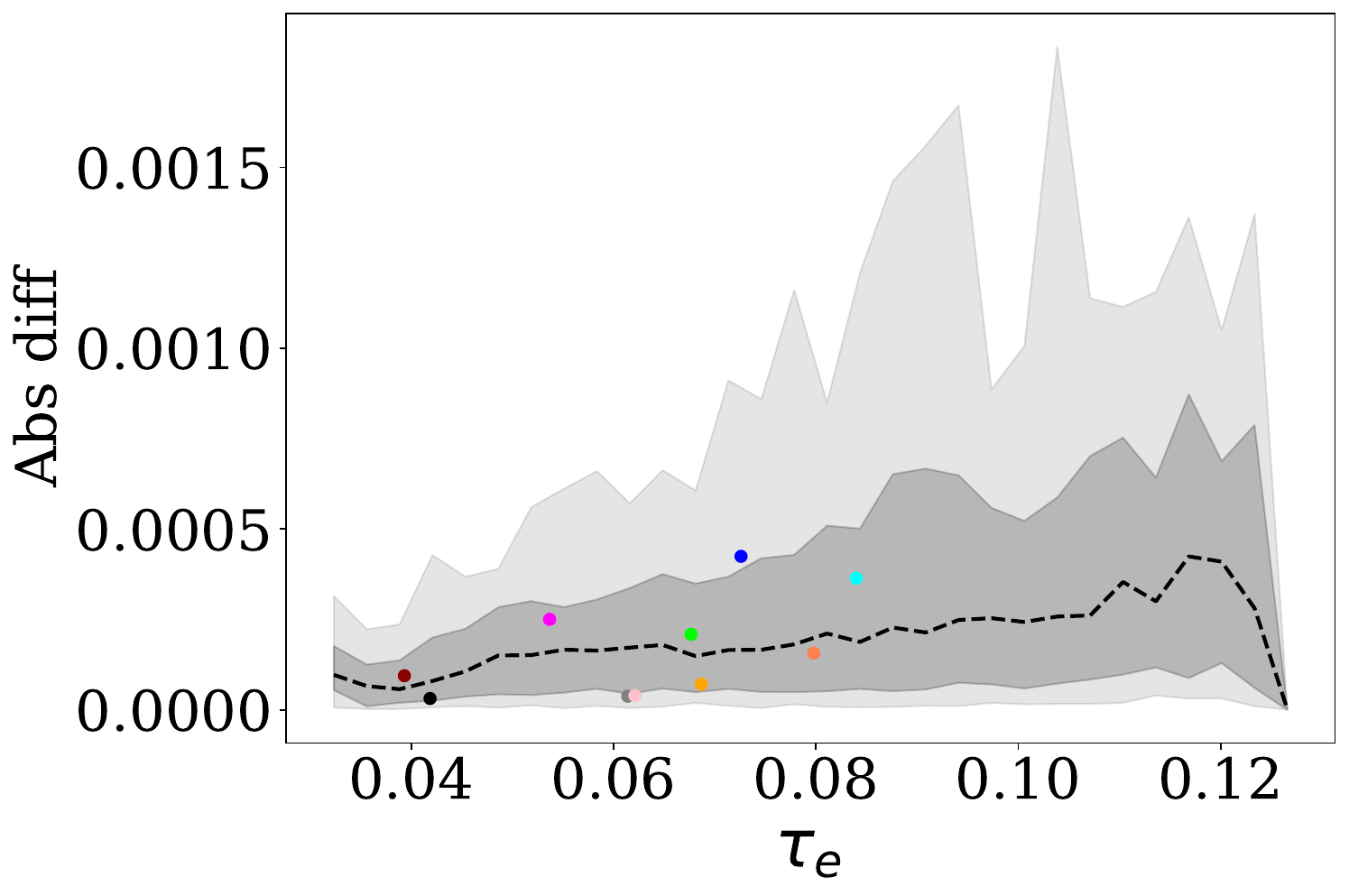}
    \end{subfigure}
    \caption{True (solid) and emulated (dashed) outputs from {\tt 21cmEMUv3} for ten random test set samples. From top to bottom, left to right: 1D 21-cm PS at $k=0.1$ Mpc$^{-1}$, mean 21-cm brightness temperature, neutral hydrogen fraction, mean spin temperature (all as a function of redshift), UV LF at $z=6$, and CMB optical depth. Bottom sub-panels show absolute differences between true and emulated quantities; individual sample errors are shown in colour and the median over the full test set in dashed black. Dark (light) shaded regions enclose 68\% (95\%) CL. The red shaded regions for the PS are evaluated against PS means rather than individual realisations.}
    \label{fig:emuvstrue}
\end{figure*}

In Figure \ref{fig:emuvstrue}, we plot true (solid) and emulated (dashed) outputs for 10 random test set samples across all summaries. The 10 samples are identical across panels. True and emulated curves are typically indistinguishable by eye, consistent with the sub-percent accuracy reported in Table \ref{tab:network_summary}. In the 1D PS panel on the top left, we additionally show individual IC realisations (thin solid lines) alongside the mean (thick solid line), highlighting that the emulated 1D PS (dashed line) is closer to the mean than any realisation. Looking at the bottom panel, we show the 68\% and 95\% confidence limits (CLs) of the absolute difference in black when comparing the emulated PS against individual realisations from about 4k different parameters. The red shaded regions show the same when comparing the emulated PS against the mean from the 90 parameter combinations. We also show the equivalent of Figure \ref{fig:emuvstrue} but with the fractional error in Figure \ref{fig:emuvstrue_fe} of the appendix.

We summarise performance in Table \ref{tab:network_summary}, where statistics are evaluated over 4k test samples for all summaries except the 1D and 2D PS, for which we use the 90-parameter mean PS test set. 
The median (68\% CL) error is 0.45\% (0.81\%) for the log 2D PS\footnote{Note that the full error covariance matrix is available within the {\tt 21cmEMU} package.}
 and 0.73\% (2.57\%) for the log 1D PS. 
 The remaining summaries are also all emulated to sub-percent median accuracy. This is particularly noteworthy in comparison to {\tt 21cmEMUv1}, which achieved a median (68\% CL) 1D PS error of 1.4\% (3.8\%) despite being trained on a database nearly two orders of magnitude larger, on a lower-dimensional summary statistic, and on a fixed IC realisation with no sample variance. This improvement can be likely attributed to the enhanced performance of the LSTM and diffusion architectures.

\begin{table}
\centering
\caption{Performance of the emulator over the test set for all summaries. Note that the test set for the PS consists of 90 mean PS, while for all other summaries, the test set contains 4k samples.}
\label{tab:network_summary}
\begin{tabular}{lll}
\hline
Summary & \multicolumn{1}{c}{Median FE (\%)} & \multicolumn{1}{c}{68\% CL (\%)} \\ \hline
$\log_{10}\delsq (k_\perp, k_\parallel)$                          &  0.45                                  & 0.81                                \\
$\log_{10}\delsq (k)$                          &  0.73                                  & 2.57                                 \\
$\aveTb$                &  0.46                                  & 1.49                            \\
$\log_{10} \aveTs$ & 0.47 & 3.48  \\
$\avenf$               &  0.02                                  & 0.29                                \\ 
$\taue$               &  0.29                                  & 0.47                                \\
$\log \phi$              &  0.21                                  & 1.33                                \\ 
                        
\end{tabular}
\end{table}

\section{Applications to inference}
\label{sec:applications}
In this section, we apply \texttt{21cmEMUv3} on a set of inference problems. We perform all inferences with the \texttt{21cmMC}\footnote{\url{https://github.com/21cmfast/21CMMC/}} driver \citep{Greig15} that can now also be run with \texttt{21cmEMUv3} in addition to \texttt{21cmFAST}, \texttt{21cmEMUv1}, and \texttt{21cmEMUv2}. \texttt{21cmMC} now also incorporates the \texttt{nautilus} \citep{Lange23} sampler, which we find provides a good balance between posterior accuracy and efficiency (see Appendix \ref{sec:samplers} for a detailed discussion). 


\subsection{Revisiting the interpretation of HERA observations}
\label{sec:hera_h1c}

\begin{table}
\caption{Comparison of the priors used in this work (two rightmost columns) with the priors used in \citealt{Lazare23} (second column). The table only lists the parameters whose prior in this work is different from \citealt{Lazare23}. The third column is the closest prior possible to \citealt{Lazare23} within the constraints of our emulator training database. The fourth column is an informed prior based on \citealt{Xu16} and ensures continuity of the SHMR around $10^8$ M$_\odot$.}
\label{tab:shmr_priors}
\begin{tabular}{llll}
\hline
Parameter & \multicolumn{1}{c}{Lazare+24} & \multicolumn{1}{c}{Lazare-like} & \multicolumn{1}{c}{Renaissance-like} \\ \hline
$\log_{10} f_{\ast,7}$ & $ [-3.5, -1]$ & $ [-3.5, -1]$ & \begin{tabular}[c]{@{}l@{}}$\mathcal{N}(-3.3, 1)$ \\  $\in [-5, -1]$\end{tabular}  \\
$\alpha_{\ast, \textsc{mcg}}$               &  $ [-0.5, 0.5] $                                 & $\alpha_{\ast,\textsc{acg}}$          & $\alpha_{\ast,\textsc{acg}}$                     \\ 
$\log_{10} f_{\ast,10}$                          &  $ [-3 , 0]$    & $ [-2, 0.5]$                              & 
\begin{tabular}[c]{@{}l@{}}$\mathcal{N}(-1,1.1)$\\  $\in [-2.5, 0.5]$\end{tabular}                              \\
$\alpha_{\ast,\textsc{acg}}$                &  $ [-0.5, 1]$     & $ [0, 1]$                              & \begin{tabular}[c]{@{}l@{}}$\mathcal{N}(0.8,0.2)$ \\$ \in [0,1] $  \end{tabular}                        \\
$\log_{10}f_{\rm esc,7}$ & $ [-3, 0]$ & $ [-3, -1]$& $ [-3, -1]$\\
$\alpha_X$ & $ [-1, 3]$ & 1 & 1\\

\end{tabular}

\end{table}

\citealt{HERA23} (henceforth HERA23) reported the deepest upper limits on the 21-cm PS to date and interpreted them with Bayesian inference using \texttt{21cmFASTv3} \citep{Mesinger07, Mesinger11, Murray20}, where the likelihood was determined by four complementary CD/EoR observables:
\begin{enumerate}[(i)]
    \item UV luminosity functions \citep{Bouwens15, Bouwens16, Oesch18} with a Gaussian likelihood;
    \item Thomson scattering optical depth to the CMB \citep{Planck18, Qin20} which uses a Gaussian likelihood around $\tau_e = 0.0569^{+0.0081}_{-0.0086}$;
    \item upper limit $\avenf < 0.06 \pm 0.05$ on the EoR history at $z=5.9$ \citep{McGreer15} obtained with the Ly$\alpha$ dark fraction method \citep{Mesinger10, McGreer15}. The likelihood function is unity at $\avenf(z=5.9) < 0.06$, decreasing as a one-sided Gaussian for higher neutral fraction values; 
    \item HERA23 upper limits on 21-cm PS with a positive unbounded prior on systematics.
\end{enumerate}

The analysis in \citealt{HERA22a} and HERA23 reveals that current upper limits on the 21-cm PS favour early galaxies that were more X-ray luminous per unit star formation rate, compared to present-day galaxies.  The X-ray luminosity of present day star-forming galaxies is dominated by high mass X-ray binary stars (HMXBs; e.g. \citealt{Mineo12}), and the inferred luminosities from HERA23 are consistent with HMXB models and  extrapolations of empirical trends to the low metallicity regime expected in the first galaxies (e.g. \citealt{Fragos13, Brorby16, Madau17, Lehmer21, Kaur22, Sartorio23}).

The HERA23 analysis, however, only accounted for HMXBs hosted in ACGs.  As pointed out in HERA23, if the first MCGs were efficient in forming stars, the corresponding HMXBs would add an additional population contributing to the X-ray background.  
This would weaken the inferred constraints on ACG X-ray efficiency, allowing for lower values than those obtained under the single-population assumption.  
This was confirmed by 
\citealt{Lazare23} (henceforth L24), who repeated the HERA analysis using an emulator of a \texttt{21cmFASTv3} model that includes both ACGs and MCGs.  L24 assumed the same X-ray efficiency per unit SFR ($L_{\rm X<2keV}/{\rm SFR}$) for both ACGs and MCGs, finding that the additional population of HMXBs hosted by MCGs resulted in weaker limits on $L_{\rm X<2keV}/{\rm SFR}$.

However, L24 demonstrated that their conclusions were heavily dependent on their choice of priors for the unknown star formation efficiency of MCGs.  Here, we revisit these results, using high-resolution hydrodynamical simulations of MCGs to motivate physical priors on their star formation efficiency.

\begin{figure*}
\begin{subfigure}{\linewidth}
    \includegraphics[width = 0.49\columnwidth]{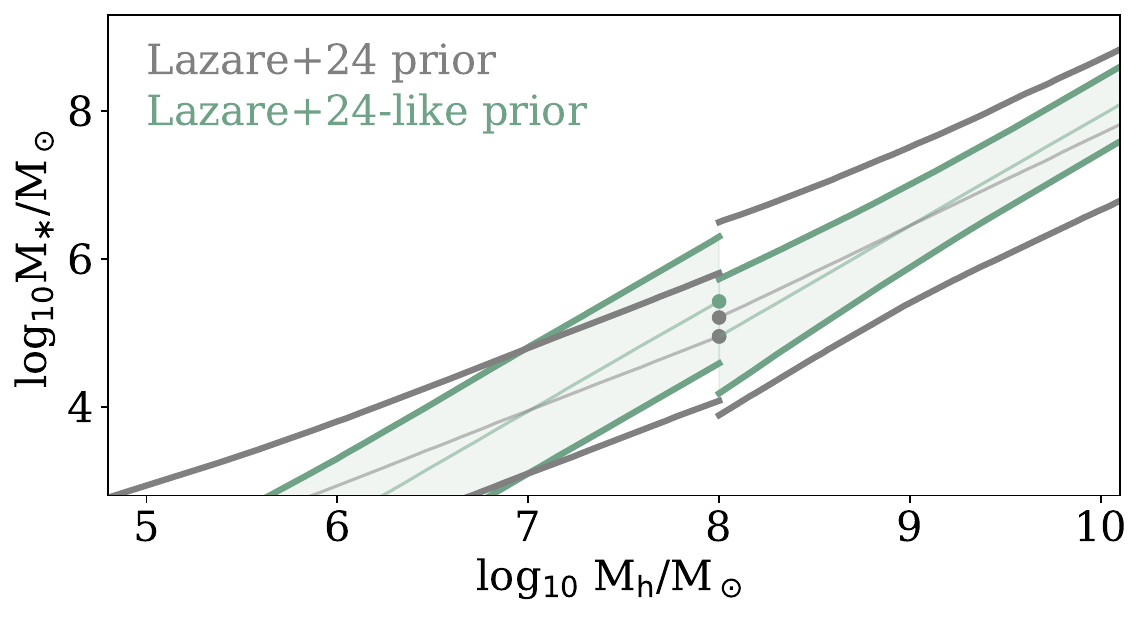} \hfill
    \includegraphics[width = 0.49\columnwidth]{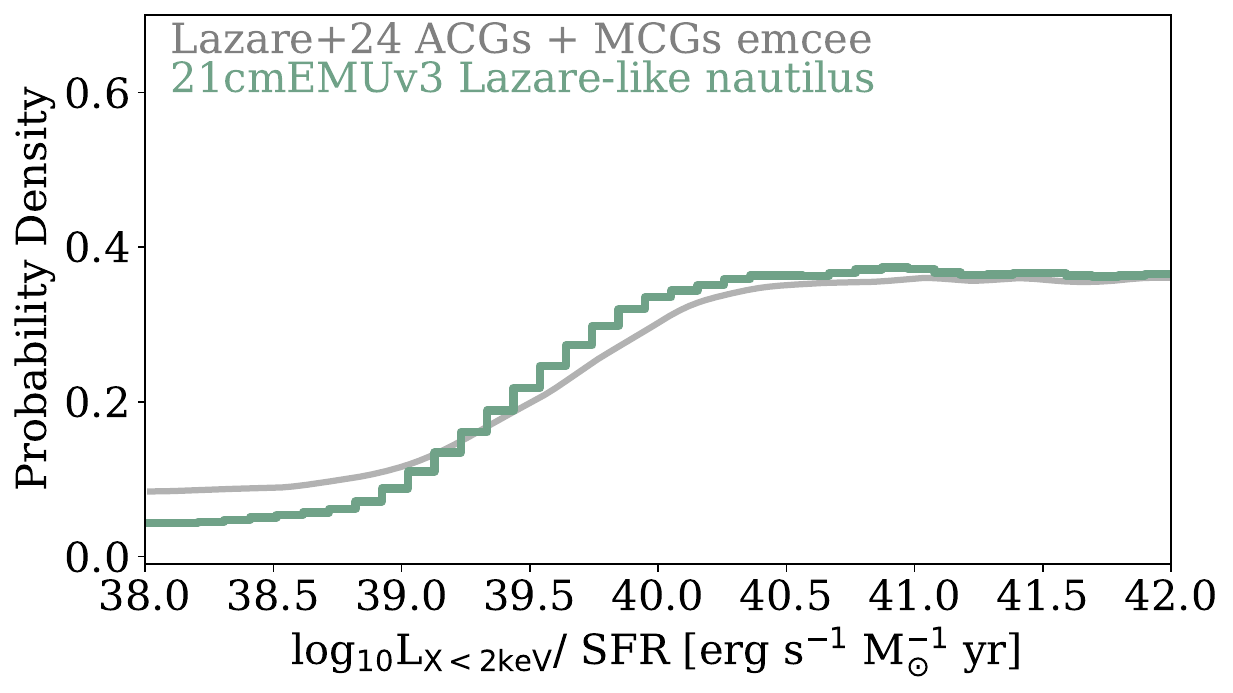}
 \end{subfigure}
    \caption{Left plot: Prior on the SHMR from L24 \citep{Lazare23} (grey). Prior used in this work that is similar to L24 but not exactly the same due to differences in the simulation database (green). Note that our L24-like prior favours lower stellar masses in comparison to L24. Right plot: Posteriors of the X-ray luminosity per unit SFR parameter for the two priors shown in the left plot.}
    \label{fig:L24_prior}
\end{figure*}

We begin by repeating  the analysis in L24 using \texttt{21cmEMUv3} with the \texttt{nautilus} sampler. Like L24, we assume $\log_{10}L_{\rm X<2keV}/{\rm SFR} \equiv \log_{10}L_{\rm X<2keV}^{\textsc{mcg}}/{\rm SFR} = \log_{10}L_{\rm X<2keV}^{\textsc{acg}}/{\rm SFR}$. Our prior differs slightly from theirs because \texttt{21cmEMUv3} was not trained on the same simulation database as their emulator. Consequently, we adopt a "L24-like" prior that is as close as possible to theirs while ensuring that it remains reasonably within the parameter space on which our emulator was trained. Table \ref{tab:shmr_priors} compares the two priors: the second column is the prior in L24, while the third column is the L24-like prior used in this work. We can compare visually the difference between the two priors in Figure \ref{fig:L24_prior}, where we plot the 68\% CL of the SHMR for their prior in grey and for our prior in green. Comparing the two we see general agreement, although our L24-like prior in green has a somewhat steeper drop in the SHMR approaching the smallest-mass MCGs with $\textrm{M}_h \sim 10^5\Msun$.

In the right panel of Figure \ref{fig:L24_prior}, we compare two inference results: the L24 result (grey) and our {\tt 21cmEMUv3} result using the L24-like prior (green)\footnote{This inference requires about 100k likelihood evaluations and takes about 20 GPU-minutes to complete.}. The latter can be obtained in two ways with {\tt 21cmEMUv3}: (i) directly from the LSTM network that emulates the 1D 21-cm PS; or (ii) by averaging the 2D PS produced by the score-based diffusion network down to 1D. We show the full corner plot including both methods in Figure \ref{fig:hera_corner} in the appendix. We find, as expected, that both approaches yield nearly identical posteriors, validating the consistency of the two emulation pathways. Comparing the green and grey posteriors, the two are qualitatively similar, though our L24-like prior provides a slightly stronger lower limit on $\log_{10}L_{\rm X<2keV}/{\rm SFR}$ due to the fact that it implies slightly less efficient star formation in $\textrm{M}_h \sim 10^5-10^6\Msun$ halos.
In both cases, the data used provide only a lower limit on  $\log_{10}L_{\rm X<2keV}/{\rm SFR}$ over our prior range.  We note however that values beyond $\log_{10}L_{\rm X<2keV}/{\rm SFR} \gtrsim  42$ start to be disfavoured by measurements of the present-day unresolved X-ray background (not considered in this work), though these constraints depend on the assumed spectral energy distribution of X-ray sources (e.g. \citealt{Mirocha25, Katz25, Dhandha25}). 

Having verified that we broadly reproduce the result from L24, we repeat the analysis with a more physically informed prior for star formation in MCGs. The informed prior choice is motivated by the Renaissance hydrodynamical simulations (\citealt{Xu16}, henceforth Xu16) and is also consistent with findings from other recent studies (e.g. \citealt{Brauer25}). The Renaissance simulations are a suite of zoom-in cosmological hydrodynamic simulations designed to model the first generation of galaxies in unprecedented detail. They simulate the formation of metal-free stars, including the process of metal enrichment due to star formation, and the effects of radiative and supernova feedback. 
Specifically, the simulations use the adaptive mesh refinement (AMR) code \textsc{Enzo} \citep{Bryan14}, coupled with its adaptive ray tracing module \textsc{Enzo+Moray} \citep{Wise11} for self-consistent on-the-fly radiative transfer. Three zoom-in regions — overdense, average, and underdense — each spanning $\sim$220--430 comoving Mpc$^3$, are evolved to final redshifts of $z = 15$, 12.5, and 8, respectively, reaching a maximum spatial resolution of 19 comoving parsecs with a dark matter particle mass of $2.9\times10^4\Msun$. The feedback prescription encompasses photo-ionization and photo-heating from H,\textsc{ii} regions, Lyman--Werner dissociation of H$_2$, and supernova feedback from both PopIII and metal-enriched stellar populations, with the resulting overpressurised H,\textsc{ii} regions and supernova-driven outflows regulating star formation through repeated cycles of gas expulsion and re-accretion in low-mass halos.

\begin{figure*}
\begin{subfigure}{\linewidth}
    \includegraphics[width = 0.49\columnwidth]{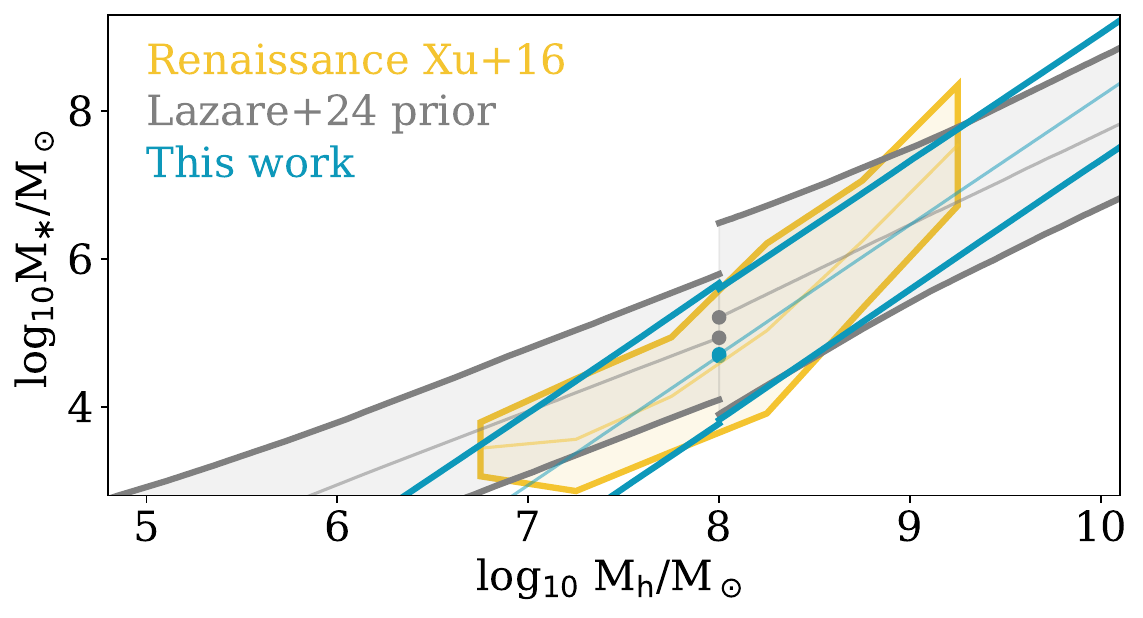} \hfill
    \includegraphics[width = 0.49\columnwidth]{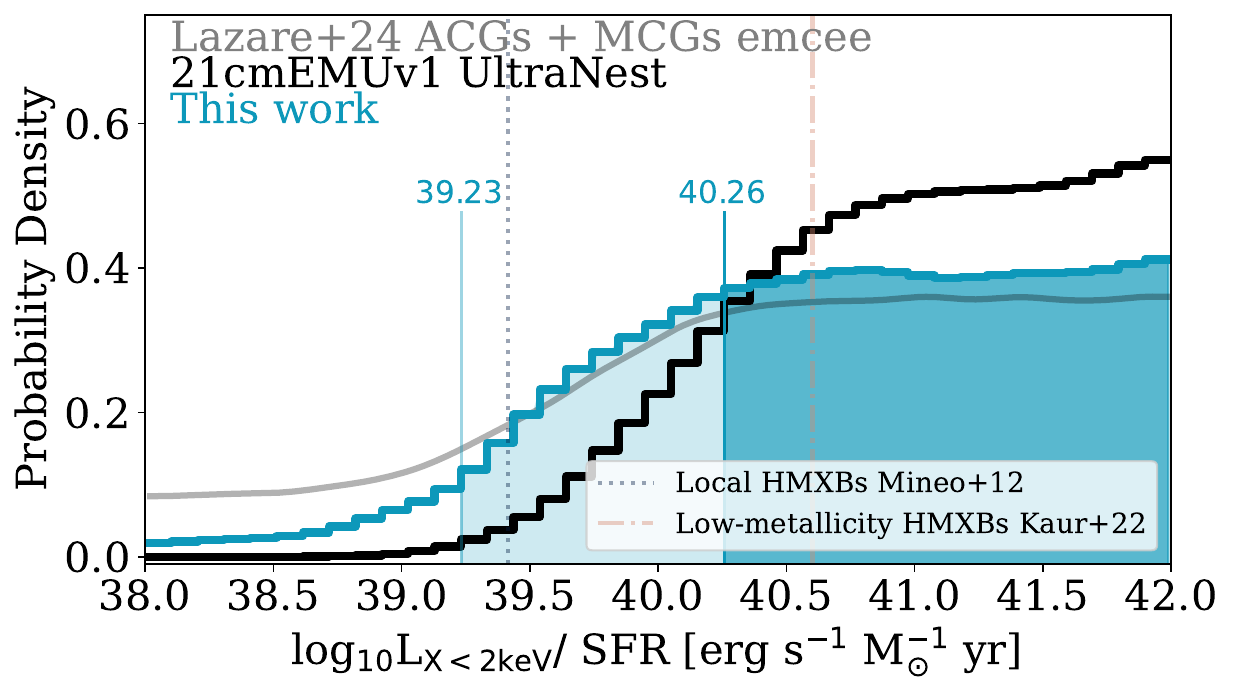}
 \end{subfigure}
    \caption{Left plot: In yellow, we show the SHMR for the Renaissance hydrodynamic simulations \citep{Xu16}.  
    We use these to motivate our prior range ({\it blue}), which favours lower stellar masses for MCGs compared to \citealt{Lazare23} ({\it gray}). Right plot: Posteriors of the X-ray luminosity per unit SFR parameter for the two priors shown in the left plot. The shaded areas correspond to 95\% highest posterior density (HPD) interval (pale) and 68\% (dark). The posterior in black is obtained from \citealt{Breitman24} with the {\tt UltraNest} sampler and {\tt 21cmEMUv1} that models only a single population of ACGs.  As expected, the posterior which uses our simulation-informed prior lies between the fiducial one of L24 and the fiducial HERA23 result that assumes only a single population of ACGs.  For reference, the vertical lines indicate mean values for local HMXBs \citep{Fragos13}, and an extrapolation of local relations to the metal-poor regime expected in the first galaxies \citep{Kaur22}.  Our new posterior is consistent with the extrapolation, and disfavours the local HMXB relation at $\sim 93\%$ CI.}
    \label{fig:informed_prior}
\end{figure*}

Xu16 examined the properties of early galaxies from populations hosted by at least a few thousand halos. In the left plot of Figure \ref{fig:informed_prior}, we show in yellow the 1$\sigma$ region of the SHMR reported by Xu16 (see their Table 2). In particular, the thin yellow curve plotted through the yellow region corresponds to the mean SHMR for a cosmic region with near-average density. The upper (lower) bounds are obtained by averaging the upper (lower) bounds of the SHMR for average, underdense, and overdense cosmic regions analysed in Xu16. The resulting yellow shaded region, therefore, spans a wide range of MCG stellar masses across diverse cosmic environments. Comparing the yellow and green 1$\sigma$ shaded regions, we see that the L24 prior favours higher stellar masses for MCGs in comparison to the Renaissance simulations.  Thus if star formation inside actual MCGs is similar to that in the Renaissance simulations, we would expect L24 to overestimate their impact on the HERA23 conclusions. 

Using the yellow shaded region for guidance, we choose a new "simulation-informed" prior shown in blue that approximates the yellow region while remaining reasonably within the parameter space on which \texttt{21cmEMUv3} was trained. We also change the ACG SHMR parameters to ensure that the priors across MCG SHMR and ACG SHMR are consistent around $10^8\Msun$. The informed prior consists of truncated Gaussian distributions. The truncation avoids samples too far outside the training range of the emulator. The parameters of the informed prior are shown in the rightmost column in Table \ref{tab:shmr_priors}.

In the right panel of Figure \ref{fig:informed_prior}, we show the 1D marginal posterior of the X-ray luminosity per unit SFR for the informed prior in blue\footnote{Note that we perform the inference twice: once with the 1D PS from the LSTM and another time with the 2D PS from the diffusion model that we then average down to 1D. We confirm that both approaches produce a nearly identical posterior, as expected. For reference, the results shown in the figures are obtained with the 1D PS LSTM emulator.}, compared to the L24 posterior in grey. Compared to L24, the informed prior more strongly disfavours the lowest values of X-ray luminosity per unit SFR. This can be understood by comparing the blue and grey 1$\sigma$ shaded regions of the priors in the left panel: the L24 prior favours higher MCG stellar masses and in turn higher SFRs.   Having a higher SFR in MCGs means that the IGM heating detected by HERA can be accomplished with lower values of X-ray luminosities per unit SFR. For reference, the vertical lines indicate the mean X-ray luminosity for local high-mass X-ray binaries (HMXBs) \citep{Fragos13} (grey dotted) and a low-metallicity extrapolation of local relations expected for the first galaxies \citep{Kaur22} (red dash-dotted). Our posterior is consistent with this low-metallicity extrapolation and disfavours the local HMXB relation at $\sim 93\%$ credible interval (CI). This re-analysis thus provides the state-of-the-art interpretation of current 21-cm observations, grounded in theoretical predictions from hydrodynamical simulations.

\subsection{SKA forecasts with different antenna layouts}
\label{sec:ska}
In this section, we use \emuiii\ to forecast CD/EoR constraints achievable with the 1D and 2D 21-cm PS from the upcoming Square Kilometre Array (SKA)\footnote{\url{https://www.skao.int/en}}. SKA-Low, the low-frequency component of the SKA relevant for 21-cm CD/EoR science, is currently under construction in Australia. Each station comprises 256 dipole antennas, covering the frequency range $50-220$ MHz — corresponding to redshifts $z \sim 5.5-27$ for the redshifted 21-cm line. Construction is proceeding in staged Array Assembly (AA) increments. Here we consider three layouts representative of different deployment stages as shown in Figure \ref{fig:ska_layouts}: (i) the fully deployed AA4 with all 512 stations; (ii) AA$^\ast$ with 307 stations; and (iii) a proposed layout P1 with 257 stations\footnote{P1 station layout obtained from Florent Mertens (private communication), based on engineering constraints provided by SKAO.}, accounting for the Fall 2025 announcement of the deferral of the construction of 50 SKA-Low stations due to budget constraints. This forecast will allow us to assess how much constraining power is lost by the deferral of 50 stations in comparison to AA$^\ast$ and full deployment. The deferred stations are predominantly located in the outskirts of the core, preferentially reducing long-baseline density while leaving the short-baseline density at the centre of the core, which drives CD/EoR sensitivity, largely intact. While the impact of the deferral on CD/EoR sensitivity is expected to be minimal, in practice, the loss of long-baseline density affects calibration, which will likely impact CD/EoR science in ways not captured by our idealised forecast, as we discuss below.

\begin{table}[ht]
\caption{Mock SKA observation fiducial parameters.}
\label{tab:mock_params}
\centering
\begin{tabular}{ll}
Parameter            & Values \\ \hline
$\log_{10} f_{*,10}$       & -1.2         \\
$\alpha_{\ast}$   & 0.5        \\
$t_{\ast}$         & 0.55  \\
$\log_{10} f_{\rm esc, 10}$     & -1.3    \\
$\alpha_{\rm esc}$     & 0.0     \\
$\log_{10} f_{\ast,7}$ &  -2.5  \\
$\log_{10} f_{\rm esc, 7}$ &-1.5\\
$\log_{10}L_{\rm X<2keV}^{\textsc{acg}}/{\rm SFR}$ & 40.5\\
$\log_{10}L_{\rm X<2keV}^{\textsc{mcg}}/{\rm SFR}$ & 41.5 \\
$E_{\rm 0}$ & 500 \\
$\sigma_8$ & 0.8118\\
\end{tabular}

\end{table}

To simulate a realistic observation, we select a parameter set from the emulator test set as the fiducial mock cosmological signal. The fiducial parameters, shown in Table \ref{tab:mock_params}, are consistent with current observational constraints such as UV LFs, and are representative of values commonly adopted in the literature (e.g.,\citealt{Qin20, Munoz22}). We modify our sensitivity simulator {\tt 21cmSense} \citep{Murray24} to interface with the SKA package {\tt ska-ost-array-config}\footnote{\url{https://gitlab.com/ska-telescope/ost/ska-ost-array-config}} in order to obtain the most up-to-date SKA antenna layouts. We then use {\tt 21cmSense} to forecast sensitivities for the three layouts. The SKA forecasting pipeline is made publicly available in the {\tt 21cmSense} documentation\footnote{\url{https://21cmsense.readthedocs.io/en/latest/tutorials/SKA_forecast.html}}. 

\begin{figure*}
\begin{subfigure}{\linewidth}
    \includegraphics[width = 0.49\columnwidth]{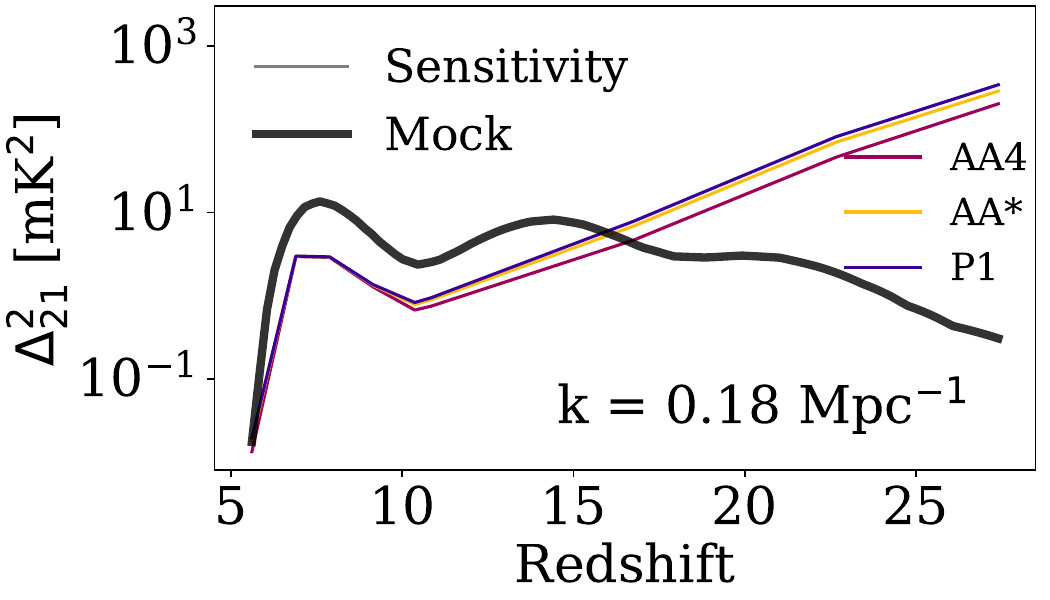} \hfill
    \includegraphics[width = 0.49\columnwidth]{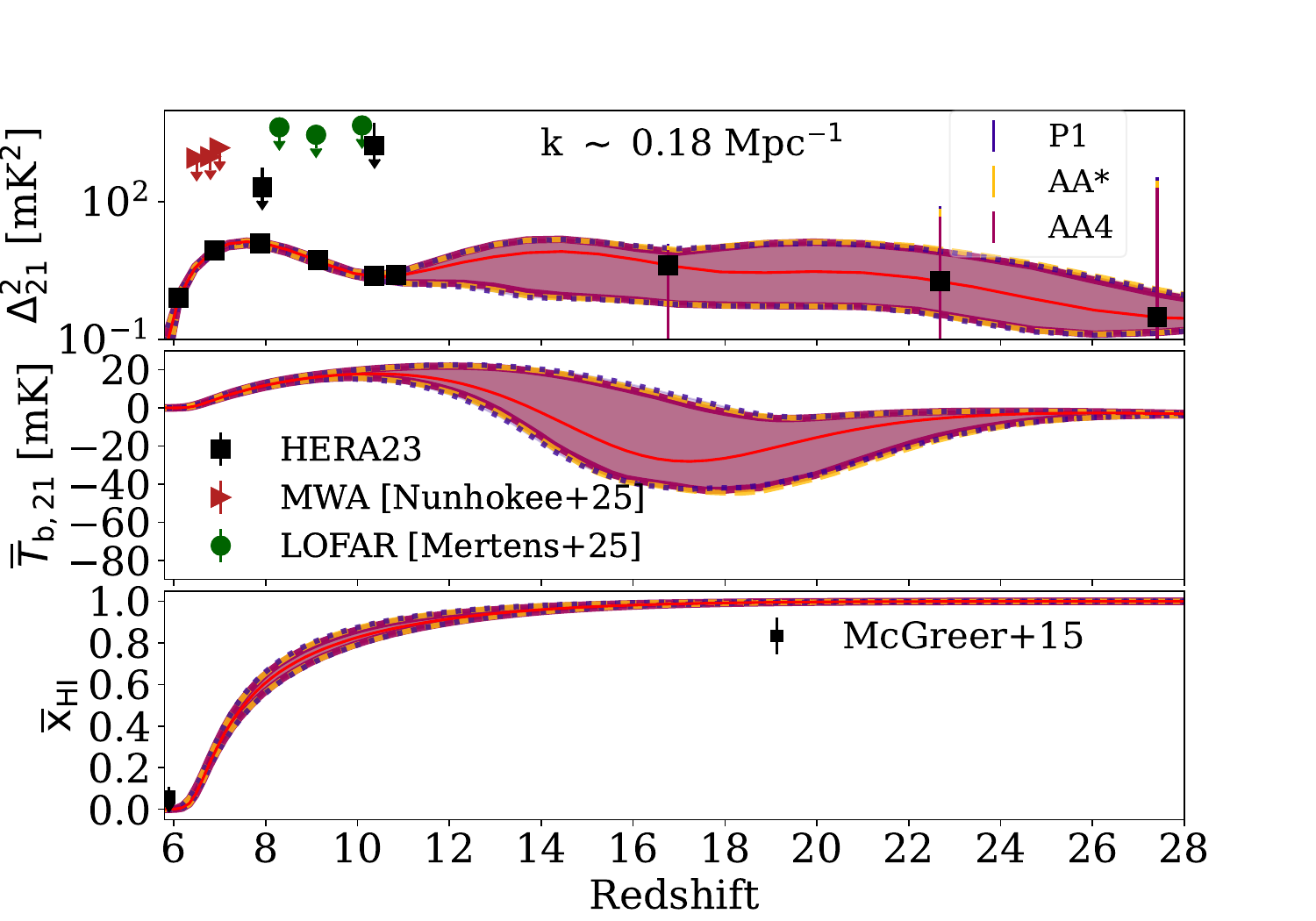}
 \end{subfigure}
    \caption{
    Left panel: Mock SKA observation of the 21-cm 1D PS at wavemode $k\sim 0.18$ Mpc$^{-1}$  (black solid), and the corresponding thermal noise sensitivity for the AA4 (purple), AA$^\ast$ (yellow), and proposed deferral P1 (blue) antenna layouts simulated with {\tt 21cmSense}. Right panel: Marginalised 95\% credible intervals on the 1D 21-cm PS (top), global 21-cm signal (middle), and neutral fraction (bottom) as a function of redshift, for the three layouts. The ground truth from the mock is shown in red in each panel.}
    \label{fig:ska_inf}
\end{figure*}

In the left panel of Figure~\ref{fig:ska_inf}, we show the mock 1D PS as a function of redshift at $k \sim 0.18$ Mpc$^{-1}$ together with the corresponding SKA thermal noise sensitivity for the three antenna layouts. At the most sensitive wavenumber and redshift in the forecast ($k \sim 0.44$ Mpc$^{-1}$, $z \sim 7.9$), the 21-cm PS is detected at $\sim 14\sigma$, $\sim 13\sigma$, and $\sim 12\sigma$ for AA4, AA$^\ast$, and P1, respectively — a $\sim 7\%$ sensitivity reduction per deployment stage at this frequency. At other redshifts, however, the reduction between successive layouts can reach $\sim 20\%$, highlighting that the impact of the 50-station deferral is frequency-dependent and most pronounced away from the peak sensitivity.

We use \emuiii\ with the {\tt nautilus} sampler to perform 11-parameter inference on the simulated mock 21-cm PS observation. The inference setup follows that of Section~\ref{sec:hera_h1c} with the Renaissance-like prior, with the sole modification that the HERA upper limits are replaced by the SKA 21-cm PS mock detections, adopting a Gaussian likelihood. The resulting posteriors on derived summaries are shown in the right panel of Figure~\ref{fig:ska_inf}. The full corner plot is provided in Figure~\ref{fig:ska_corner} in Appendix \ref{sec:cornerplots}.

We find that the additional baseline density of the AA4 layout provides only marginal improvement in constraining power relative to AA$^\ast$, suggesting that AA$^\ast$ already captures most of the astrophysical information encoded in the 21-cm PS for this fiducial model. The EoR history is well recovered across all three layouts, with 95\% CI on the neutral fraction narrowing from $\Delta\bar{x}_\mathrm{HI} = 0.098$ (P1) to $0.085$ (AA$^\ast$) to $0.077$ (AA4) around the midpoint of reionisation. The global 21-cm signal is similarly well constrained. Most notably, the Thomson scattering optical depth $\tau_e$ is recovered to $\sim$5--6 times the precision of the \citet{Planck18} measurement across all three layouts, demonstrating that even the most conservative deployment stage would represent a dramatic improvement over CMB-based constraints on the reionisation history (though see \citealt{Qin25} for state-of-the-art constrains on the EoR history from the Lyman alpha forest). We note that these constraints represent a relatively conservative estimate, as they are obtained by varying all 10 astrophysical parameters and $\sigma_8$ simultaneously. Simpler models with fewer free parameters yield considerably tighter bounds. For instance, forecasts using {\tt 21cmEMUv1}, which varies 9 parameters under a less flexible galaxy model, recover $\Delta\bar{x}_\mathrm{HI} \approx 0.03$ and constrain $\tau_e$ to $\sim$20 times the precision of \citet{Planck18}. The constraining power reported here should therefore be interpreted in the context of the model complexity and inference dimensionality.

The forecasts presented here are idealised and should be interpreted with care. While our sensitivity calculations do assume a moderate foreground avoidance model excluding Fourier modes within the foreground wedge, a region in cylindrical $k$-space contaminated by spectrally smooth foregrounds (e.g., \citealt{Morales12, Parsons12, Liu14a}), plus a $0.1\,h\,\text{Mpc}^{-1}$ buffer (e.g. see \citealt{HERA22}), they do not account for residual systematics that currently limit real observations.  \citet{Mertens25} illustrate their importance, demonstrating that a more contaminated LOFAR field can yield better 21-cm PS upper limits than a cleaner one, owing largely to the greater maturity of its calibration and foreground subtraction pipeline. This shows that the achievable sensitivity strongly depends on the accuracy of sky modelling and calibration rather than thermal noise alone. Since long baselines are essential for building accurate sky models and constraining gain solutions, the reduced long-baseline density of the AA$^\ast$ and P1 layouts may increase residual foreground contamination in ways not captured by our forecasts. The actual impact of the station deferral on CD/EoR science is therefore likely more significant than suggested here. We further note that science cases relying heavily on long baselines, such as very long baseline interferometry (VLBI), are expected to be considerably more affected by the deferral than CD/EoR science. Overall, despite these caveats, our forecasts suggest that the scientific return of SKA-Low for CD/EoR science remains strong even under the deferred deployment schedule.

\section{Conclusion} \label{sec:conclusion}

We have presented {\tt 21cmEMUv3}, an emulator of {\tt 21cmFASTv3} modelling two galaxy populations: atomically and molecularly cooling galaxies, conditioned on eleven astrophysical and cosmological parameters. {\tt 21cmEMUv3} jointly predicts seven summary observables using a score-based diffusion model for the cylindrical 21-cm PS, LSTM networks for the five time-evolving summaries, and a feed-forward network for the optical depth to reionization, achieving sub-percent median fractional error across all outputs. Notably, the 2D PS is emulated at a better accuracy than the 1D PS from {\tt 21cmEMUv1}, despite being a significantly more complex summary statistic and being trained on a database nearly two orders of magnitude smaller.

We apply our emulator to re-interpret the HERA23 measured upper limits on the 21cm PS, together with complimentary observations.  HERA23 discovered a new source of IGM heating at $z\gtrsim10$, likely provided by HMXBs in the first galaxies.  The fiducial HERA23 analysis only considered HMXBs in ACGs.  As quantified in \citealt{Lazare23}, the inclusion of an additional population of MCG-hosted HMXBs would weaken the implied constraints on the X-ray luminosity per unit star formation, simply by increasing the total star formation density of the Universe.  

We revisit this claim, using for the first time a simulation-informed prior for star formation in MCGs.  Our prior is motivated by the Renaissance hydrodynamical simulations \citep{Xu16} that favour less efficient star formation inside MCGs in comparison to the fiducial prior in \citealt{Lazare23}. 
As expected, our resulting constraints on the X-ray luminosity per unit star formation are in between those of L24 and HERA23. 
These results provide a simulation-informed interpretation of current 21-cm upper limits that grounds MCG priors in theoretical predictions from state-of-the-art hydrodynamical simulations, thus trading dependence on an arbitrary prior choice for dependence on the specific star-formation and feedback prescriptions of the Renaissance simulations. Our posterior is consistent with low-metallicity extrapolations of empirical HMXB trends, and disfavours the local HMXB relation at $\sim 93\%$ C.I.

Finally, we showcase {\tt 21cmEMUv3} by producing SKA parameter forecasts from mock 21-cm PS observations in $\sim$1 GPU-hour, over 6 orders of magnitude faster than with direct simulation on CPUs. We consider three SKA-Low layouts representative of different deployment stages: the fully deployed AA4 (512 stations), AA$^\ast$ (307 stations), and a proposed P1 layout (257 stations) reflecting the Fall 2025 announcement of the deferral of 50 stations due to budget constraints. The 21-cm PS is forecast to be detected at $\sim14\sigma$, $\sim13\sigma$, and $\sim12\sigma$ for AA4, AA$^\ast$, and P1, respectively, at the most sensitive wavemode and redshift ($k\sim0.44$ Mpc$^{-1}$, $z\sim7.9$), with sensitivity reductions between successive layouts of $\sim7\%$ at peak sensitivity and up to $\sim20\%$ at other redshifts and wavemodes. Notably, AA4 provides only marginal improvement over AA$^\ast$, suggesting that the latter already captures most of the available astrophysical information for this fiducial model. A joint 11-parameter inference with the SKA mock 1D PS recovers the EoR history with a 95\% CI of $\Delta\bar{x}_\mathrm{HI} = 0.098, 0.085$, and $0.077$ for P1, AA$^\ast$, and AA4, and constrains $\tau_e$ to $\sim$5--6 times the precision of \citet{Planck18}. 
 We caution that these forecasts are idealised, as we do not account for residual systematics outside of the foreground wedge.
 It is precisely the inclusion of the long-baselines with successive deployments which will allow us to improve calibration and mitigate systematics.  Overall, however, our forecasts suggest that even the most conservative deployment stage will deliver transformative constraints on the CD/EoR astrophysics and cosmology, and that the impact of the station deferral on CD/EoR science is modest, suggesting that the scientific return of SKA-Low remains strong even under the deferred deployment schedule.

\section*{Acknowledgements}
D.B. thanks Nicholas Kern for useful discussion about the project, and Ely Kovetz for commenting on the manuscript. We gratefully acknowledge computational resources of the Center for High Performance Computing (CHPC) at SNS.
A.M. and R.T. acknowledge support from the Italian Ministry of Universities and Research (MUR) through
the PRIN project "Optimal inference from radio images of the epoch of reioniza-tion".
S.M. has received funding from the European Union’s Horizon 2020 research and innovation programme under the Marie Skłodowska-Curie grant agreement No. 101067043.
\section*{Data Availability}

\texttt{21cmEMU} is on a publicly accessible github repository, as well as available for install as a Python package using \texttt{pip}.



\bibliographystyle{aa}
\bibliography{refs}




\appendix
\section{Emulator performance}
\label{sec:emu_errs}

\begin{figure*}
\begin{subfigure}{\linewidth}
    \includegraphics[width = 0.49\columnwidth]{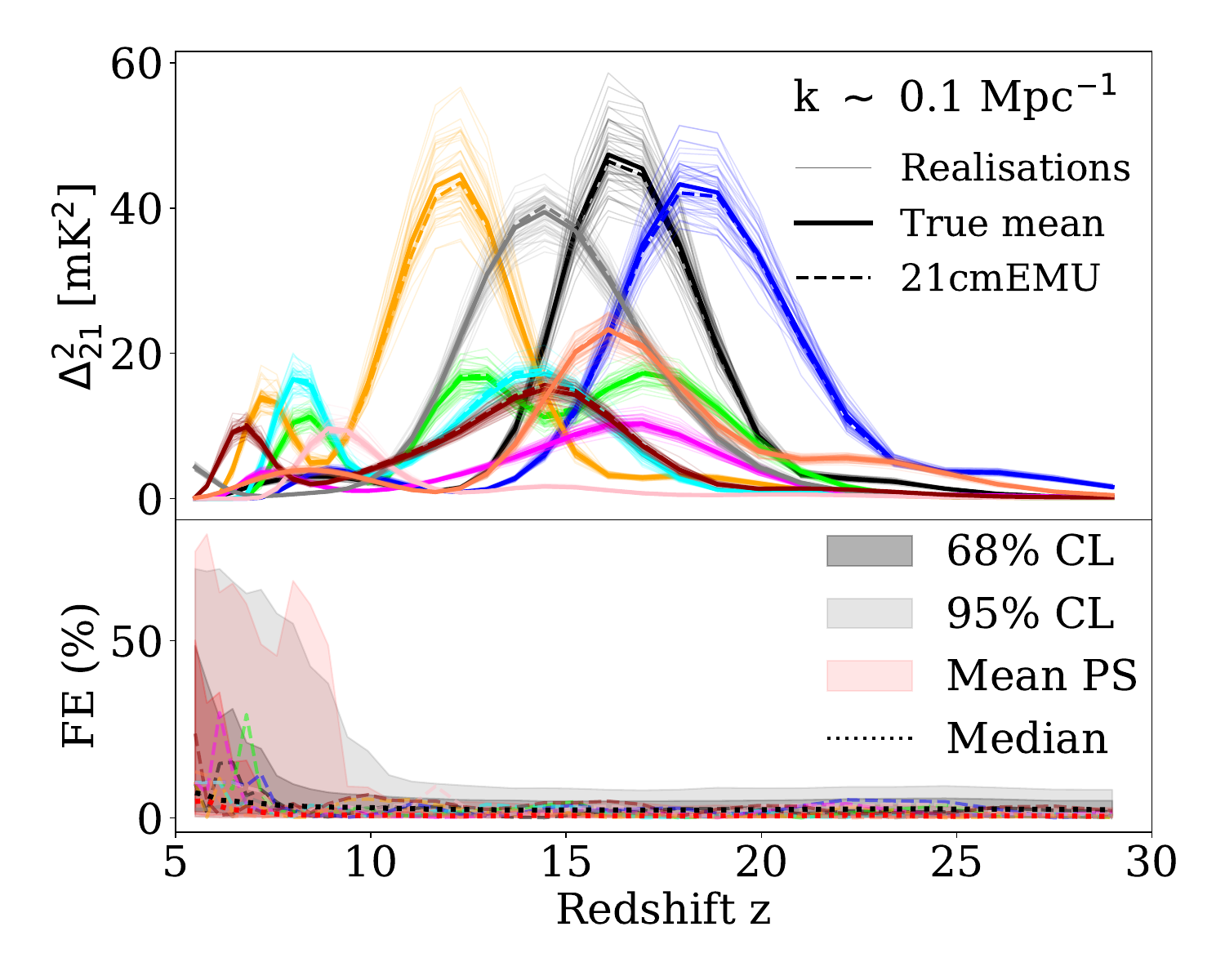} \hfill
    \includegraphics[width = 0.49\columnwidth]{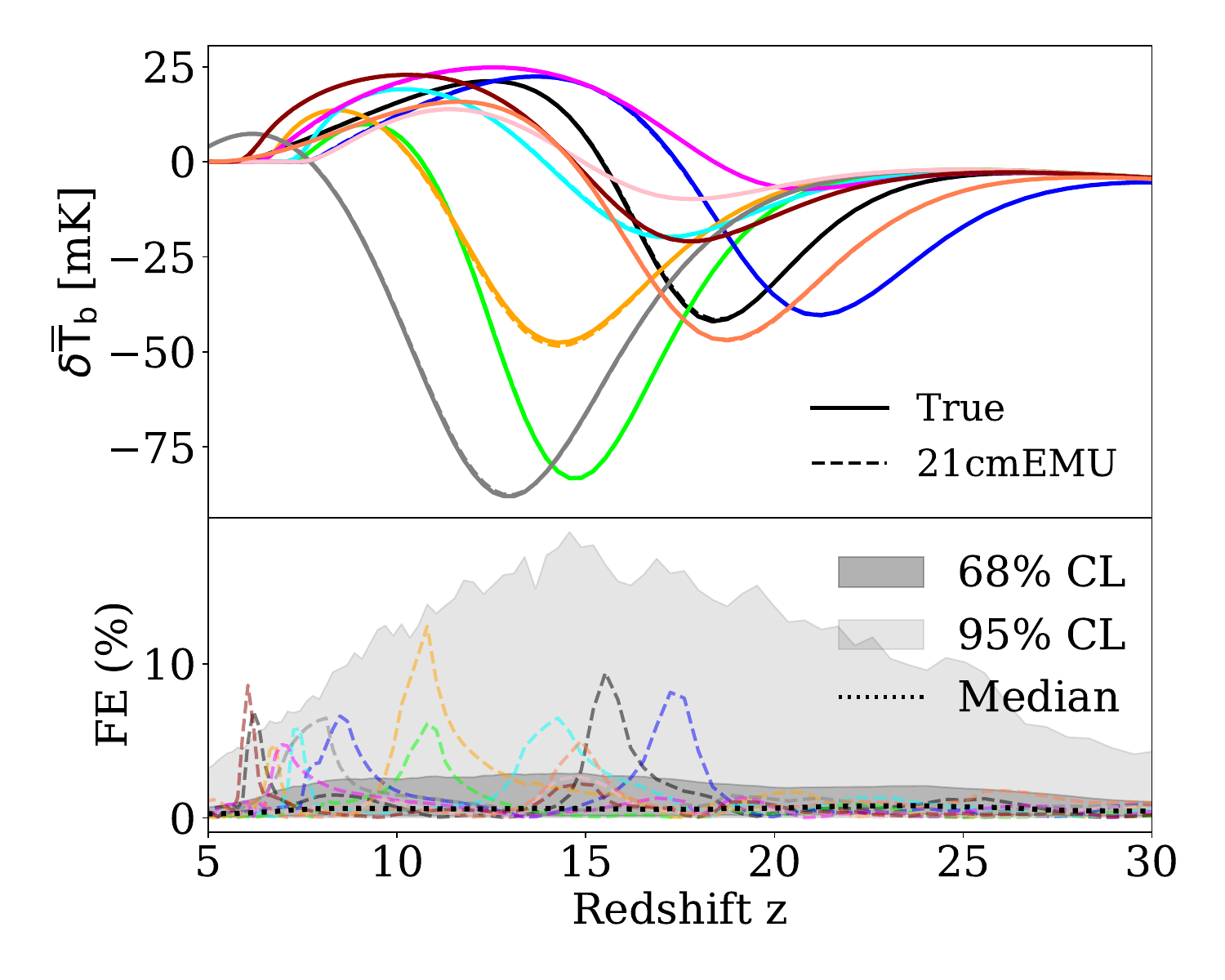} \\

    \includegraphics[width = 0.49\columnwidth]{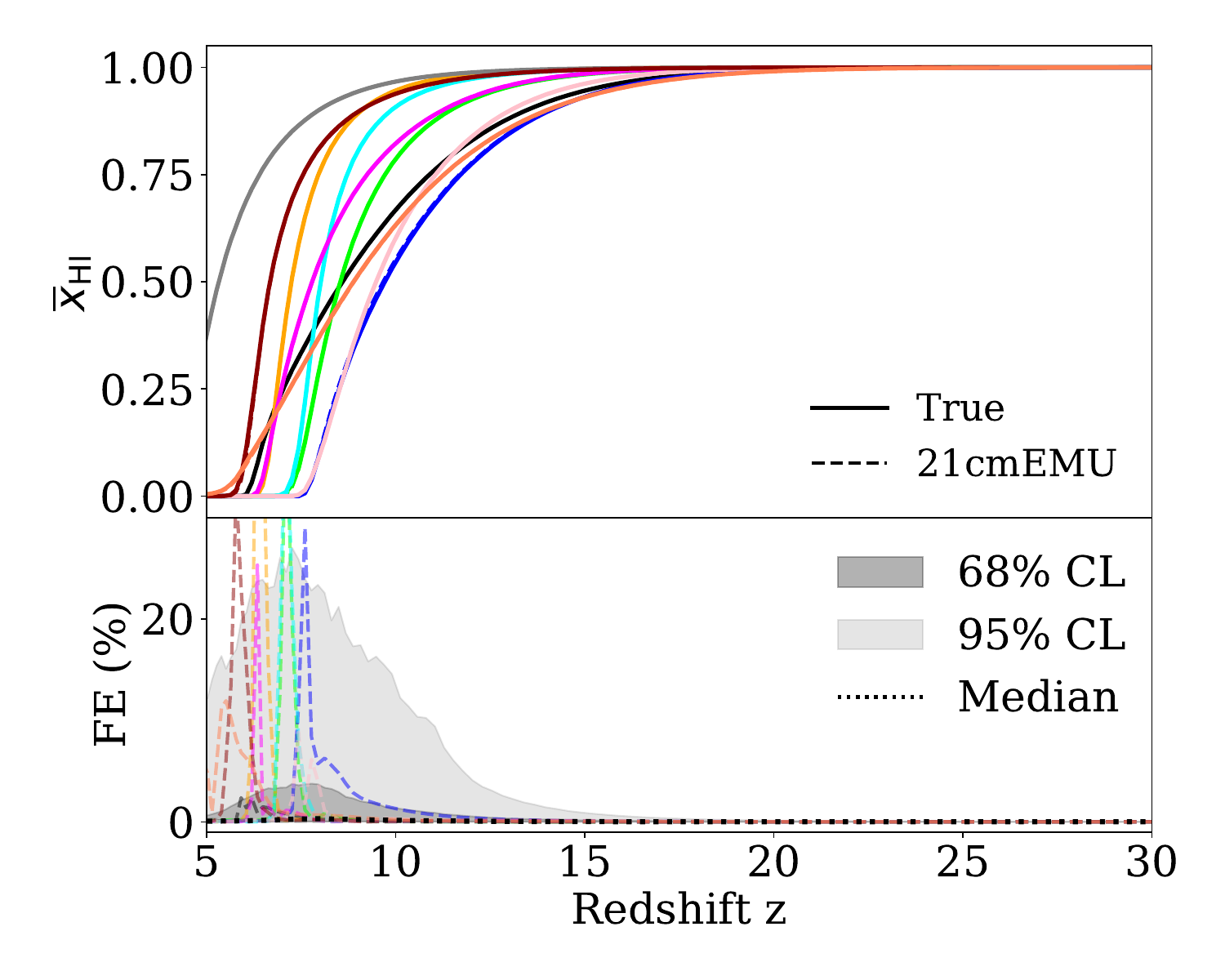}\hfill
    \includegraphics[width = 0.49\columnwidth]{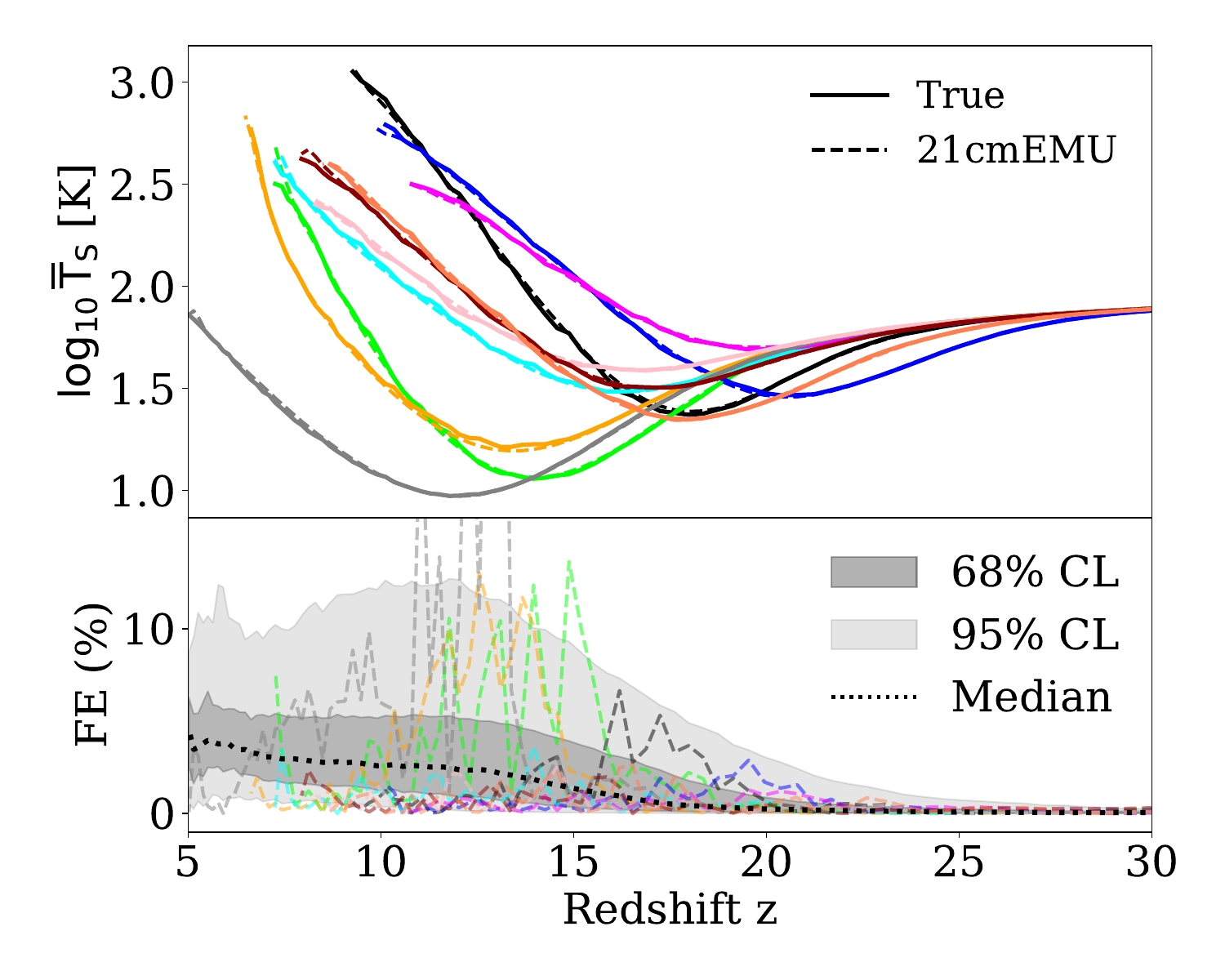} \\
    \includegraphics[width = 0.49\columnwidth]{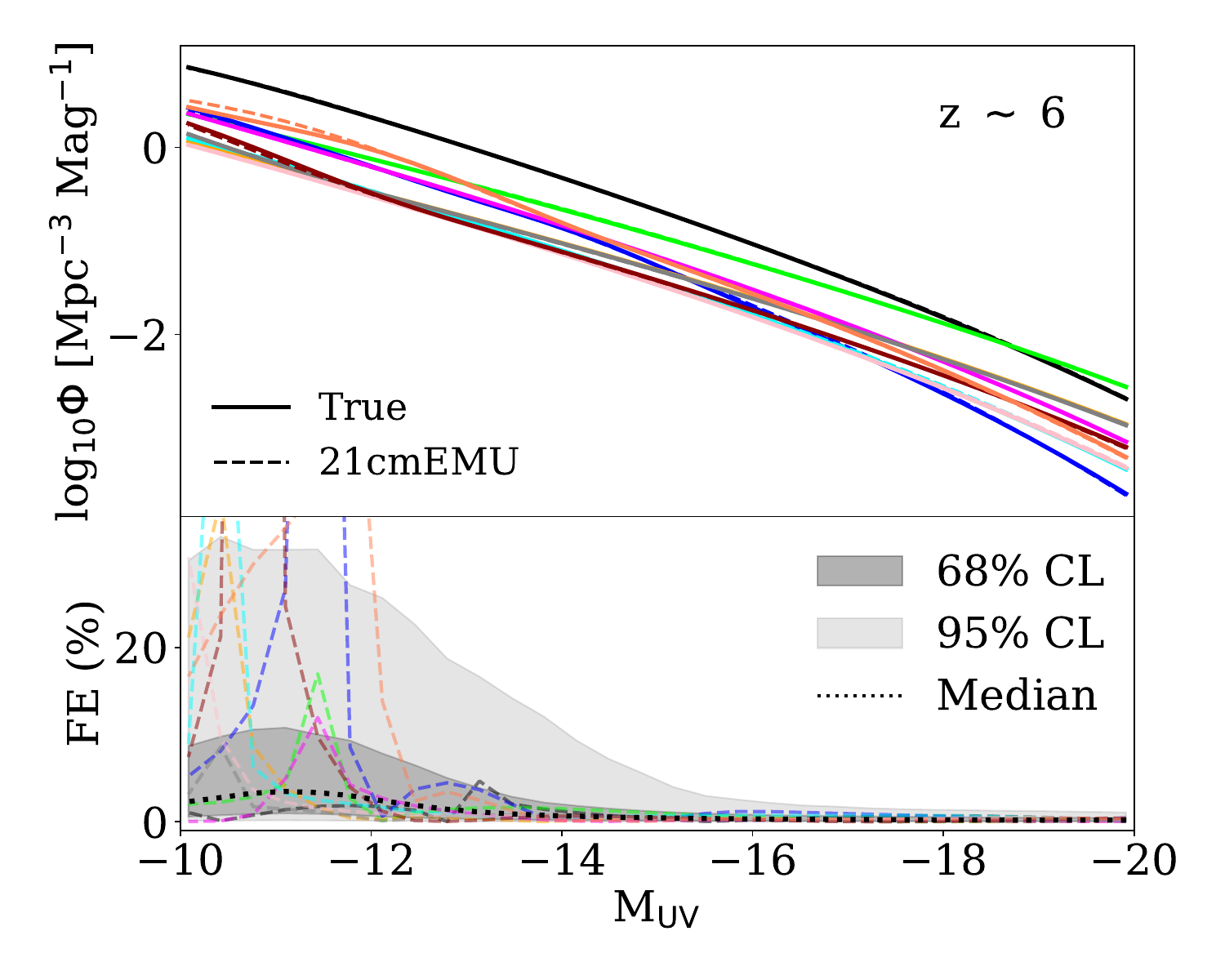} \hfill
    \includegraphics[width = 0.49\columnwidth]{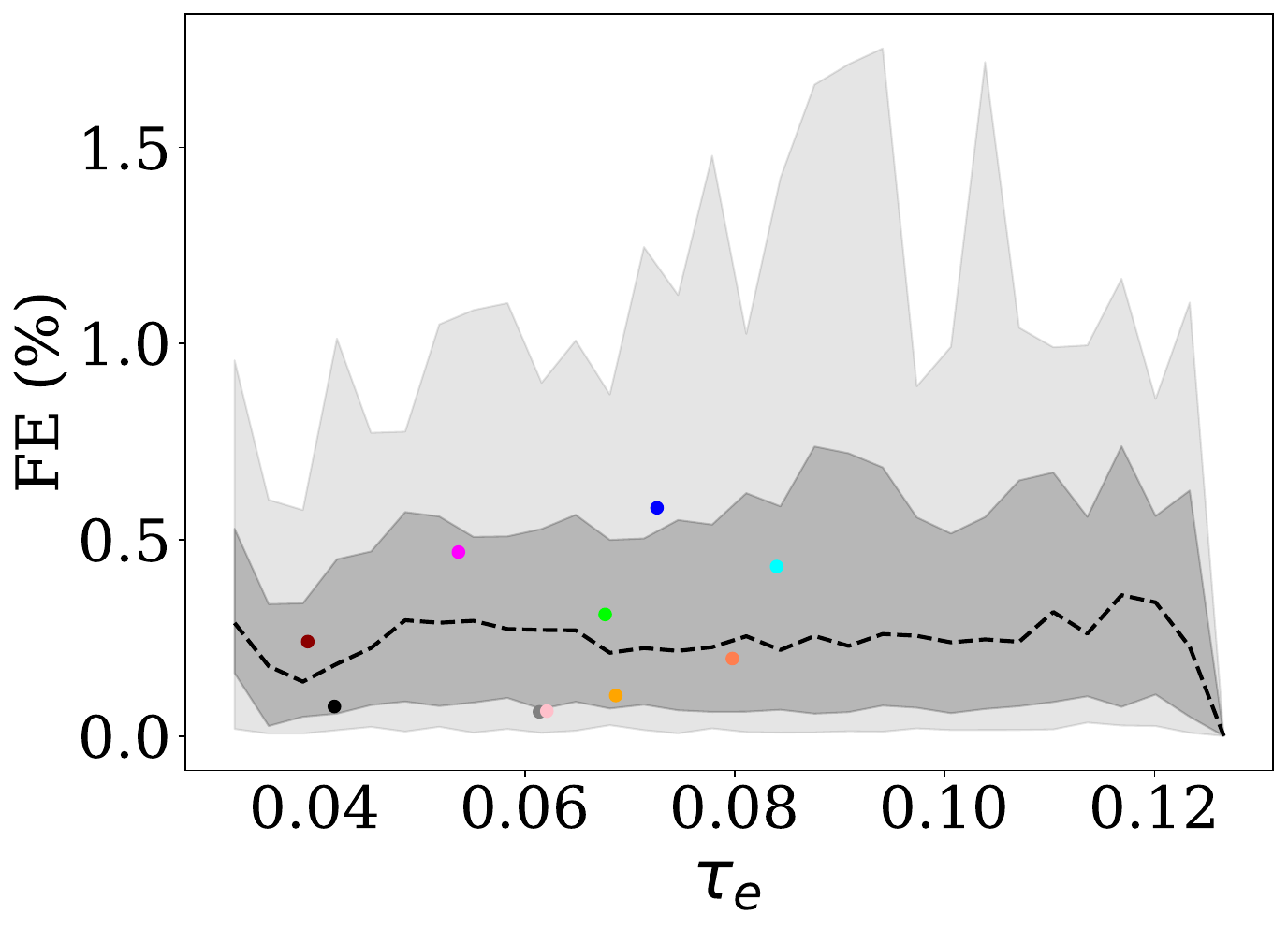}
    \end{subfigure}
    \caption{True (solid) and emulated (dashed) outputs from {\tt 21cmEMUv3} for ten random test set samples. From top to bottom, left to right: 1D 21-cm PS at $k=0.1$ Mpc$^{-1}$, mean 21-cm brightness temperature, neutral hydrogen fraction, mean spin temperature (all as a function of redshift), UV LF at $z=6$, and CMB optical depth. Bottom sub-panels show fractional errors between true and emulated quantities; individual sample errors are shown in colour and the median over the full test set in dashed black. Dark (light) shaded regions enclose 68\% (95\%) CL. The red shaded regions for the PS are evaluated against PS means rather than individual realisations.}
    \label{fig:emuvstrue_fe}
\end{figure*}

\begin{figure*}
    \centering
    \includegraphics[width=0.9\linewidth]{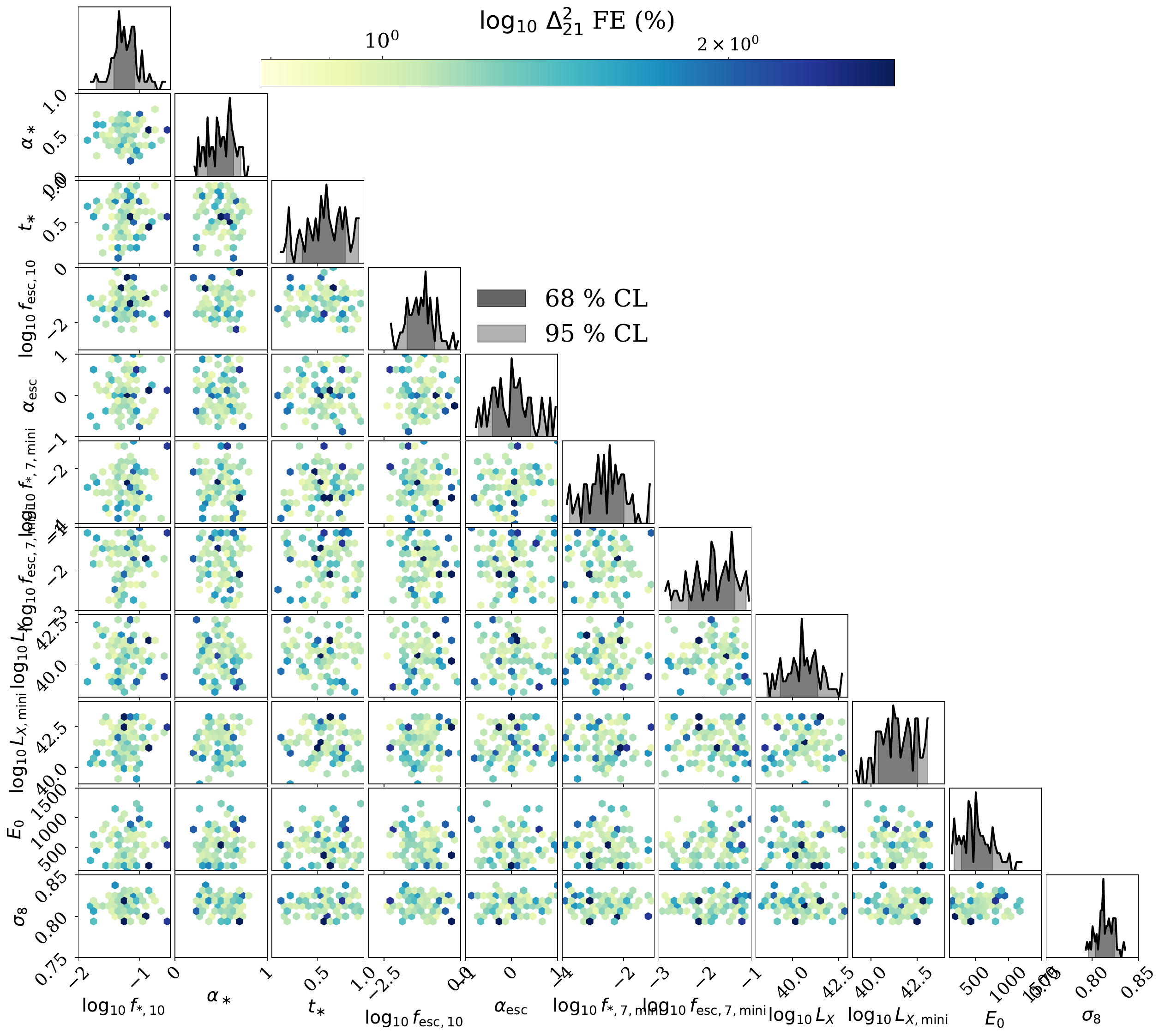}
    \caption{Median linear 2D PS fractional error as a function of parameters for the 90-parameter test set. }
    \label{fig:FE_corner}
\end{figure*}

This appendix provides additional performance plots for {\tt 21cmEMUv3}, supplementing the accuracy metrics discussed in Section~\ref{sec:emu}.
Figure~\ref{fig:emuvstrue_fe} shows true (solid) and corresponding emulated (dashed) outputs for ten random test set samples across all six LSTM and feed-forward network summaries: the 1D 21-cm PS, mean 21-cm brightness temperature, neutral hydrogen fraction, mean spin temperature, UV LF at $z=6$, and CMB optical depth. Individual fractional errors are shown in colour, with the median over the full test set in dotted black and 68\% and 95\% confidence limits shaded. The sub-percent median accuracy achieved across all summaries and redshifts is consistent with the results reported in the main text. Note that for the 1D PS (upper left panel), we also shade in red the FE distribution comparing emulated PS to mean PS for the 90-parameter test set.
Figure~\ref{fig:FE_corner} shows the median fractional error of the 2D PS as a function of the eleven input parameters, evaluated on the 90-parameter test set. This allows us to identify any regions of parameter space where the diffusion emulator performs less well, and confirms that the sub-percent accuracy is maintained broadly across the parameter space rather than being driven by a particularly well-sampled region.

\section{Impact of different samplers}
\label{sec:samplers}

Nested sampling (NS; \citealt{Skilling04, Skilling06}) is a popular choice for Bayesian inference, as it facilitates model comparison by efficiently calculating the Bayesian evidence.  There are several choices of public NS algorithms.  \texttt{MultiNest} \citep{Feroz09} is one of the first science grade NS samplers.  It is numerically efficient, typically requiring fewer likelihood evaluations than other algorithms for 21-cm inferences.
However, in \citealt{Breitman24}, we found \texttt{MultiNest} to be inaccurate close to prior edges (see also \citealt{Hee16, Dittman24}). We found that   \texttt{UltraNest} \citep{Buchner21} yielded more accurate posteriors close to prior edges; however, the inferences in \citealt{Breitman24} were  numerically demanding, requiring about 12M likelihood evaluations and taking about 6 GPU hrs with 5k livepoints. Since \texttt{21cmEMUv3} includes a score-based diffusion model that is about 3 orders of magnitude slower than the MLP in \texttt{21cmEMUv1}\footnote{Note that if not needed, the score-based diffusion component can be turned off in \texttt{21cmEMUv3}, in which case it is roughly as fast as \texttt{21cmEMUv1}.}, here we use \texttt{nautilus}: a NS algorithm that balances accuracy and efficiency.
The choice of \texttt{nautilus} over other NS algorithms was motivated by its performance in \citealt{Lange23} and especially \citealt{Albert25} which presents an in-depth comparison of many MCMC and NS algorithms on various problems in physics and cosmology. 
\begin{table}[]
\caption{Each inference uses 2k livepoints and the same prior and likelihood. The number of effective samples $n_{\rm eff}$ is the number of posterior samples considered independent from one another and is calculated with Kisch's formula in Eq. \ref{eq:Kisch}. The efficiency is $n_{\rm eff} / n_\mathcal{L}$.}
\label{tab:sampler_compare}
\begin{tabular}{llll}
Algorithm                            & $n_{\rm eff}$ & $n_\mathcal{L}$ & Efficiency \\ \hline
\texttt{UltraNest} &       10k    &     2.3M        &    $0.004$        \\
\texttt{MultiNest}  &      8k     &    52k        &   0.15         \\
\texttt{nautilus}   &     10k      &    113k       &    0.09       
\end{tabular}

\end{table}

\begin{figure*}
\begin{subfigure}{\linewidth}
    \includegraphics[width = 0.49\columnwidth]{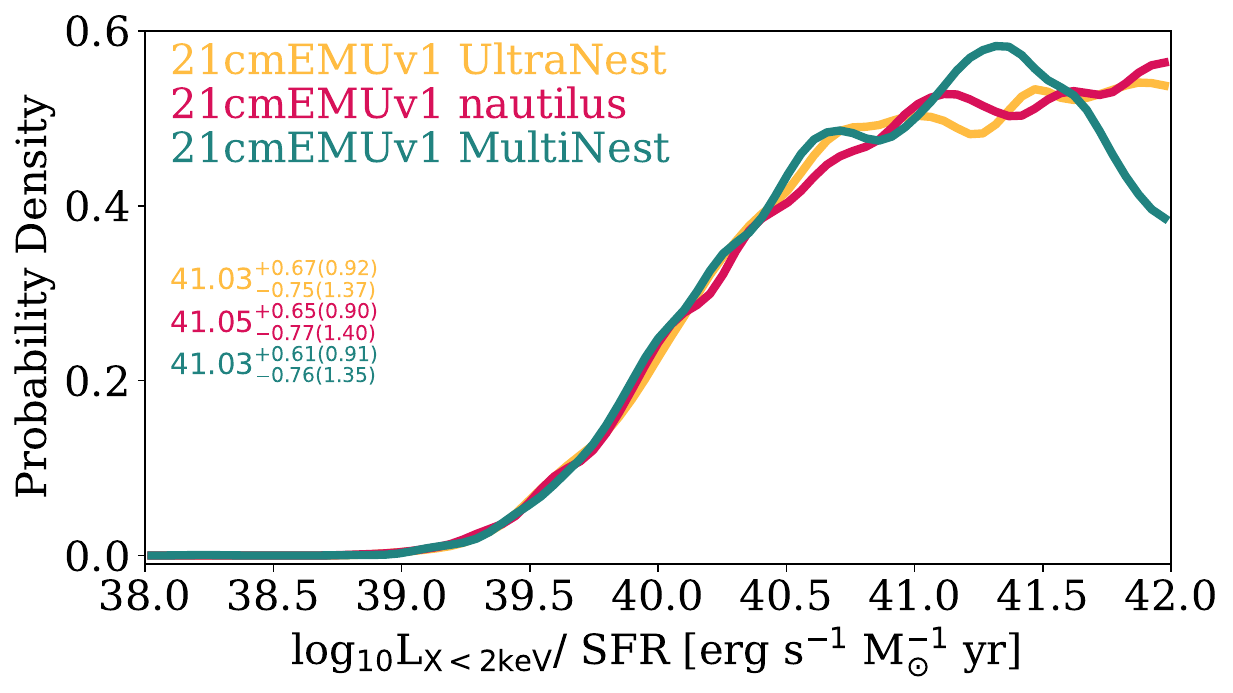} \hfill
    \includegraphics[width = 0.49\columnwidth]{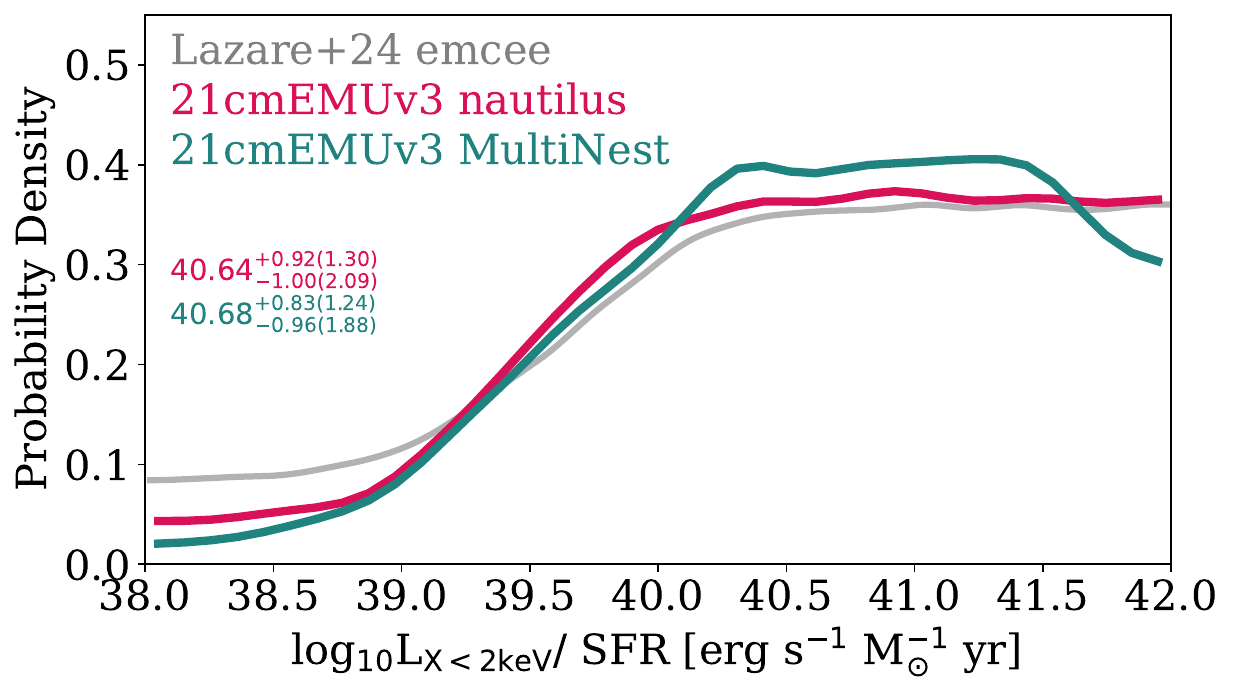}
 \end{subfigure}
    \caption{Left plot: The posterior on X-ray luminosity per unit SFR for the same inference performed with \texttt{UltraNest} (yellow), \texttt{nautilus} (red), and \texttt{MultiNest} (green) using \texttt{21cmEMUv1}. Right plot: Similar to the left plot, we are comparing inferences with different samplers. In grey, we show the result from \citealt{Lazare23} obtained with \texttt{emcee}. We show the posteriors obtained using \texttt{21cmEMUv3} with \texttt{nautilus} (red) and \texttt{MultiNest} (green) from a similar inference as the one in grey (see Section \ref{sec:hera_h1c} for more details).}
    \label{fig:sampler_comparison}
\end{figure*}

In the left panel of Figure \ref{fig:sampler_comparison}, we show the 1D marginal posterior of the X-ray luminosity per unit SFR for the same inference performed with three different samplers and 2k livepoints: \texttt{UltraNest} (yellow), \texttt{MultiNest} (green), and \texttt{nautilus} (red). To compare their efficiency, we compute the effective sample size (ESS; $n_{\rm eff}$) for each sampler using Kish's formula \citep{Kish65}:
\begin{equation} \label{eq:Kisch}
    n_{\rm eff} = \frac{\left( \sum_i w_i \right)^2}{\sum_i w_i^2},
\end{equation}
where $w_i$ is the posterior weight of sample $i$. The efficiency, defined as the ratio $n_{\rm eff} / n_\mathcal{L}$, where $n_\mathcal{L}$ is the total number of likelihood evaluations, is reported for each sampler in the right column of Table \ref{tab:sampler_compare}. We find that \texttt{nautilus} produces a posterior nearly identical to \texttt{UltraNest}, while being over an order of magnitude more efficient, making it the best choice among the three samplers for this inference problem.

\begin{figure*}
    \centering
    \includegraphics[width=0.9\linewidth]{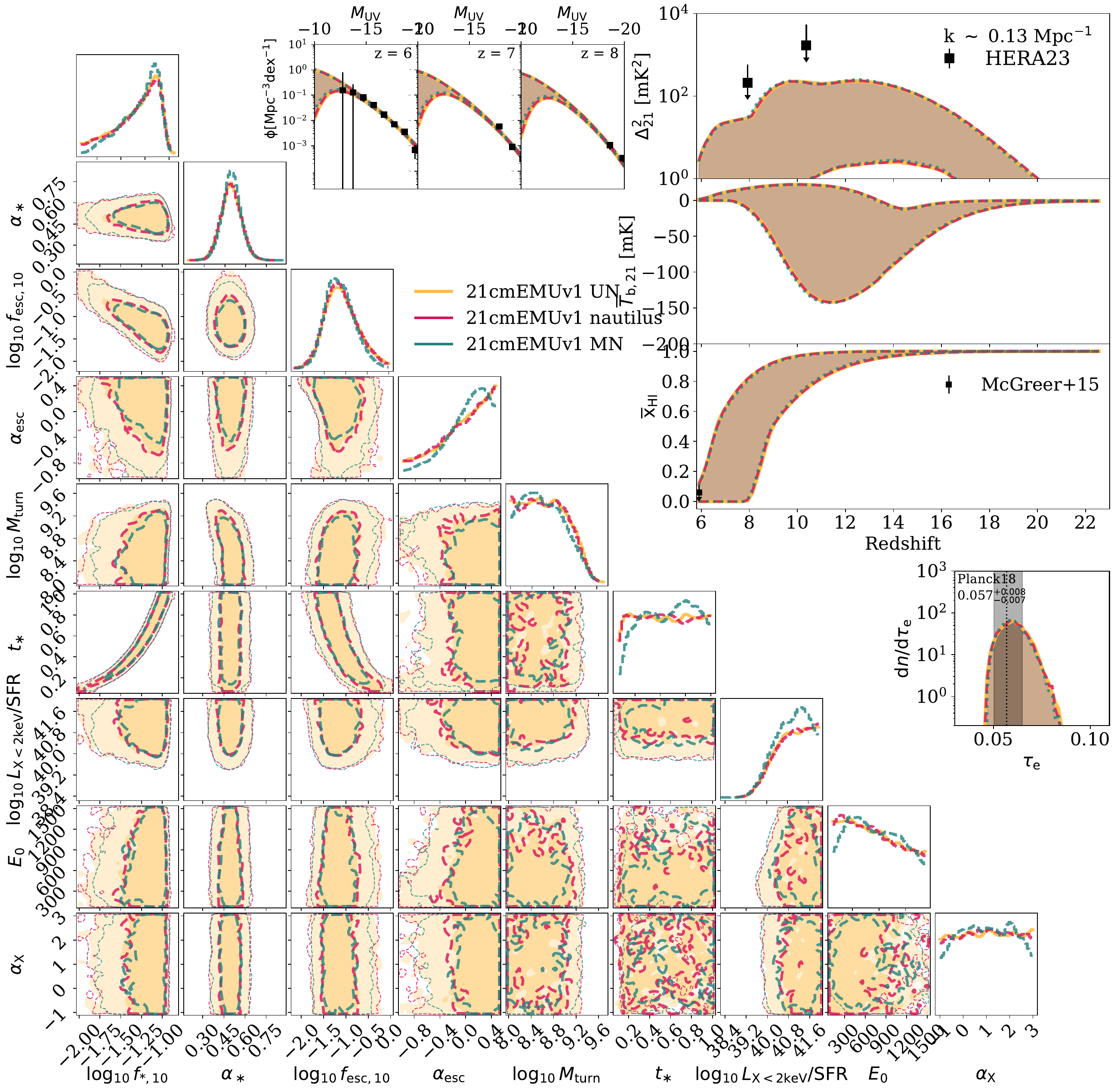}
    \caption{Comparing the three samplers with \texttt{21cmEMUv1}}
    \label{fig:samplers_corner_v1}
\end{figure*}

\begin{figure*}
    \centering
    \includegraphics[width=0.9\linewidth]{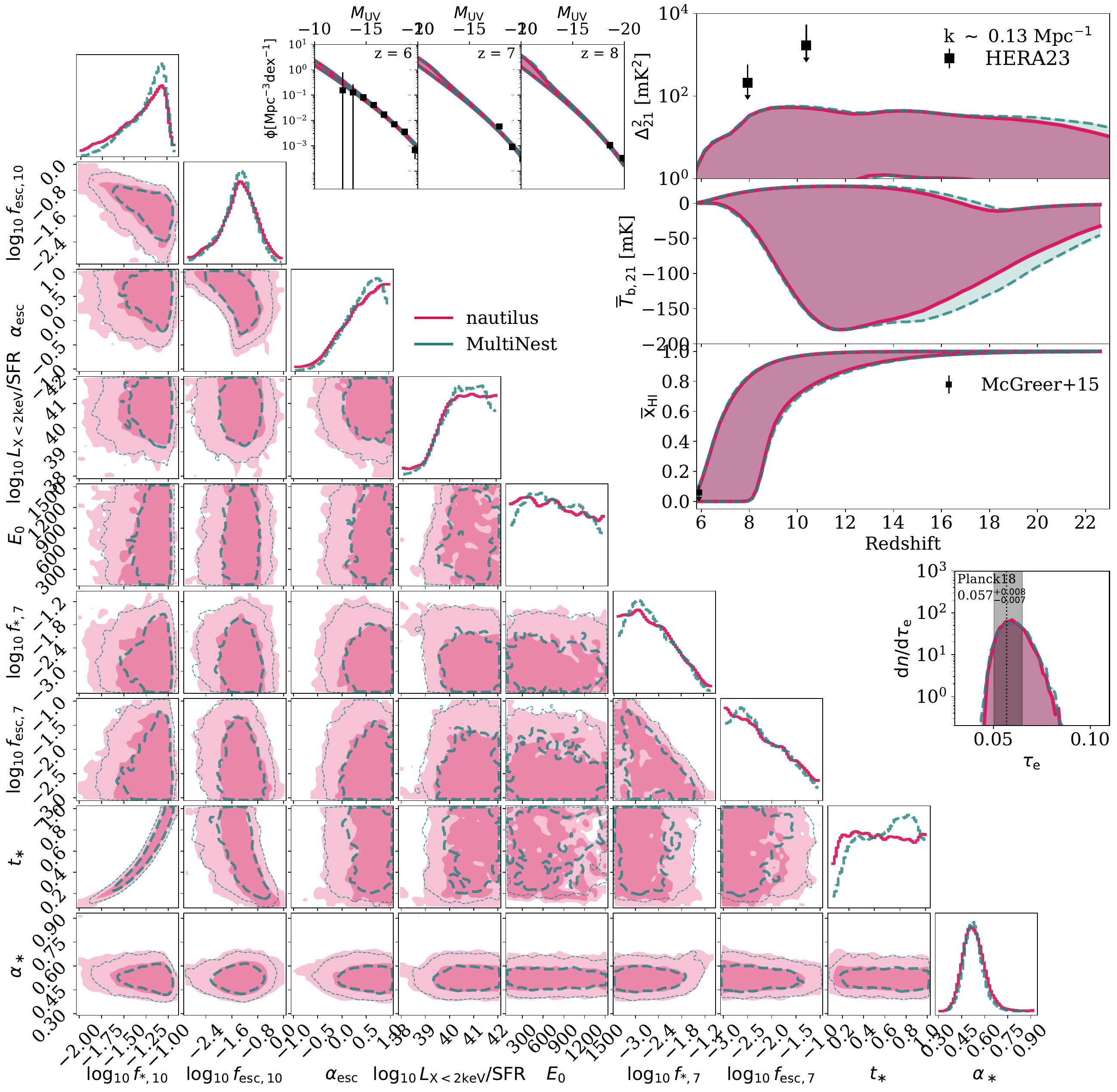}
    \caption{Comparing two samplers with \texttt{21cmEMUv3}}
    \label{fig:samplers_corner_v3}
\end{figure*}

In the right plot of Figure \ref{fig:sampler_comparison}, we perform a similar comparison to the one in the left plot, but with \texttt{21cmEMUv3} instead. Here we only compare \texttt{MultiNest} and \texttt{nautilus}. We can see the same trends in this plot as in the one on the left, where \texttt{MultiNest} tends to undersample at the prior edges. We present the full corner plots for these comparisons in Figures \ref{fig:samplers_corner_v1} and \ref{fig:samplers_corner_v3}. Based on these comparisons, we conclude that \texttt{nautilus} provides the best balance between posterior accuracy and efficiency, and we therefore use it for all the inferences in this work.


\section{Full posterior distributions}
\label{sec:cornerplots}

\begin{figure*}
    \centering
\includegraphics[width=0.4\linewidth]{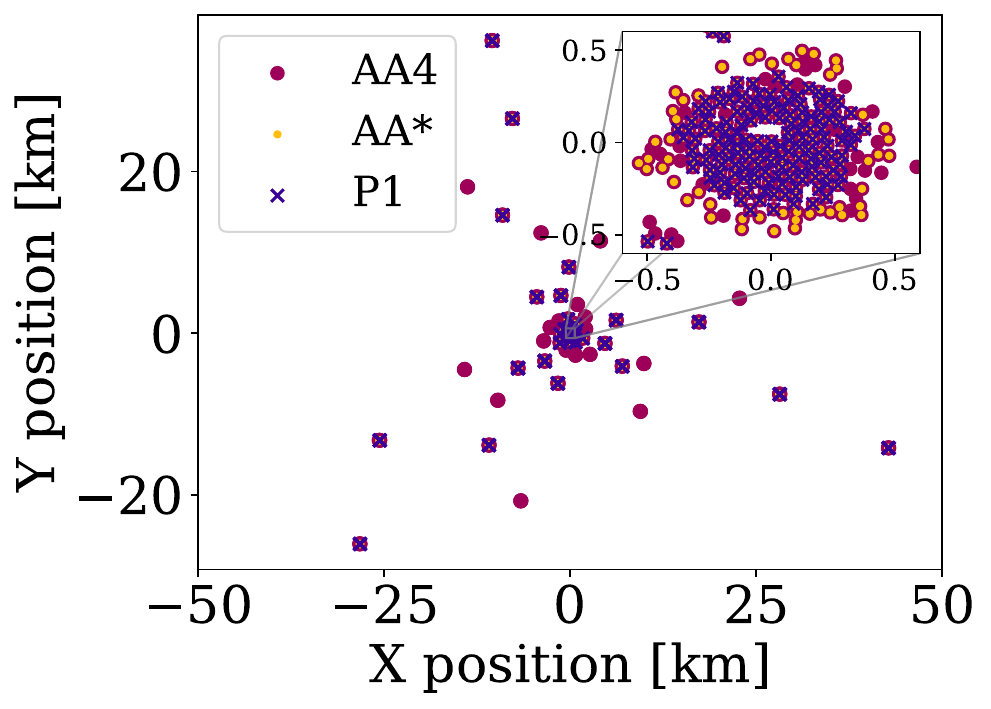}
    \caption{SKA station layouts for the three layouts considered in this work: AA4 with 512 stations (purple), AA$^\ast$ with 507 stations (yellow), and P1 with 257 stations (blue).}
    \label{fig:ska_layouts}
\end{figure*}

\begin{figure*}
    \centering
    \includegraphics[width=0.9\linewidth]{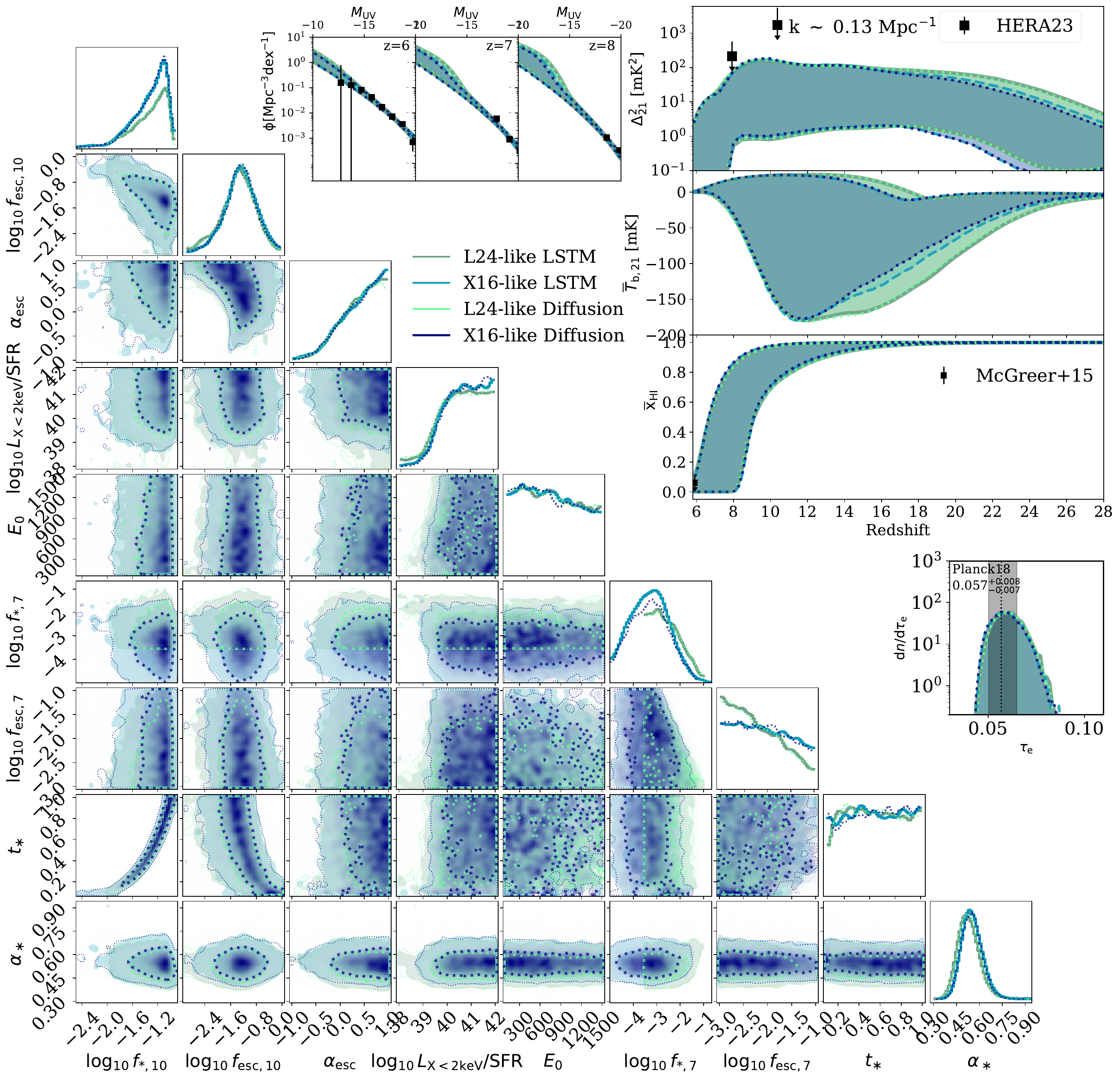}
    \caption{Comparing inferences with the Renaissance-like prior based on \citealt{Xu16} and L24-like prior. We perform the analysis in two ways: (i) with the LSTM network that directly produces the 1D PS; and (ii) with the score-based diffusion network that produces the 2D PS that we then average down to 1D. We can see that both methods produce consistent results.}
    \label{fig:hera_corner}
\end{figure*}

\begin{figure*}
    \centering
    \includegraphics[width=0.9\linewidth]{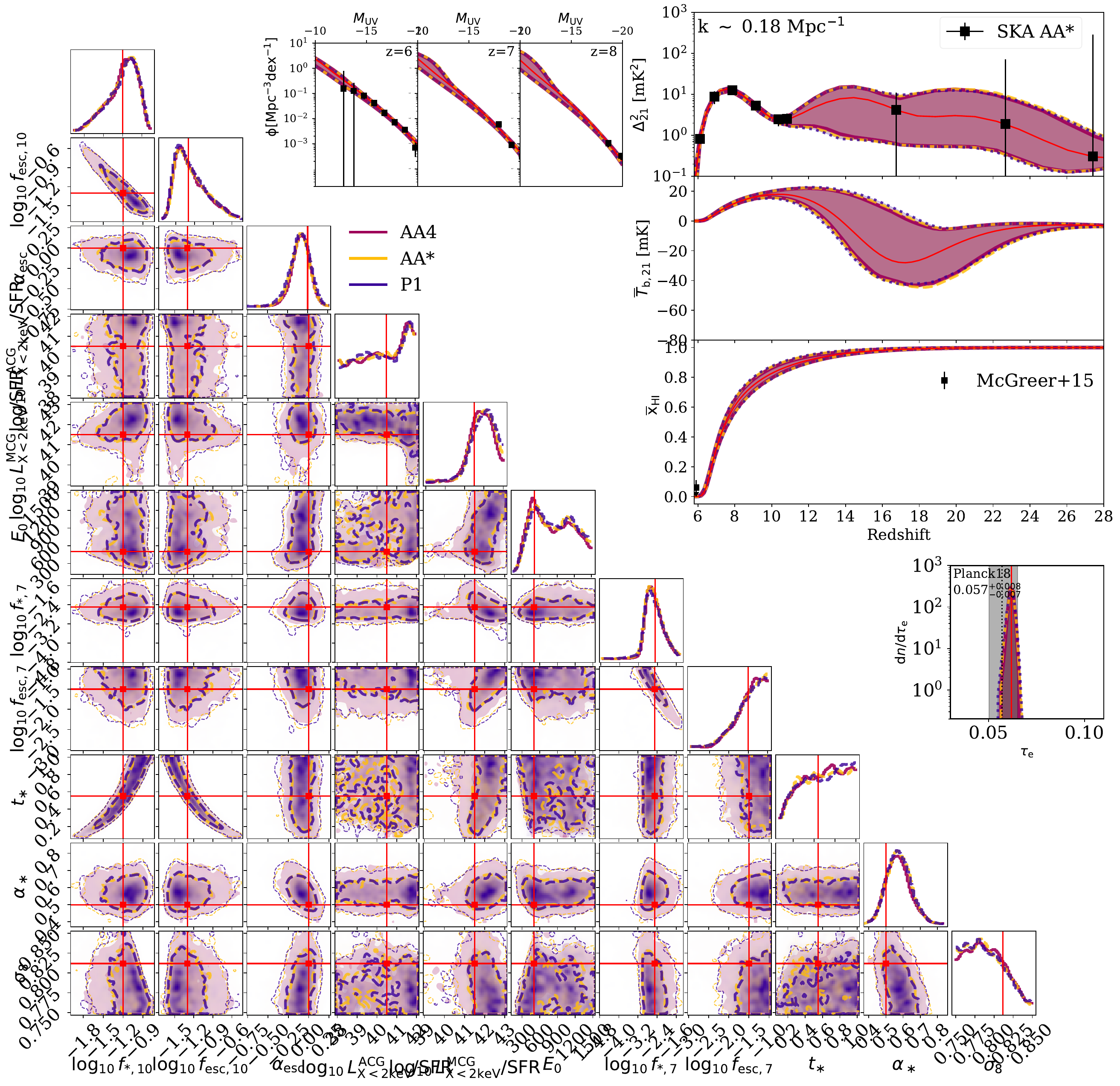}
    \caption{Full corner plot from inferences with SKA mock observations for the AA4 (purple), AA$^\ast$ (yellow), and P1 (blue) SKA layouts.}
    \label{fig:ska_corner}
\end{figure*}

In this appendix, we show the full posterior distributions for the inferences presented in the main text.
Figure~\ref{fig:ska_layouts} shows the station layouts for the three SKA-Low configurations considered in this work. Figure~\ref{fig:hera_corner} shows the full corner plot for the HERA23 re-analysis comparing the L24-like and Renaissance-like priors, with both the LSTM (1D PS) and score-based diffusion (2D PS averaged to 1D) emulation pathways. The two pathways are consistent among eachother across all eleven parameters. Figure~\ref{fig:ska_corner} shows the full corner plot for the SKA forecast inference, comparing the posterior distributions obtained with the AA4, AA$^\ast$, and P1 layouts. There is marginal improvement from AA4 over AA$^\ast$ and P1, as all three layouts yield visually similar posteriors in spite of the increasing station count.

\section{2D PS to 1D PS}

We can obtain the 1D spherically averaged 21 cm PS from the 2D cylindrical PS. The 2D PS is averaged down to 1D by weighting each 2D PS bin by the number of 3D Fourier modes it contains, $N_k$, as follows:
    \begin{align}
        \Delta^2_{21} (k_i) = \frac{1}{\sum_{k \in \mathcal{K}_i} N_k(k_\perp,k_\parallel)}\sum_{k \in \mathcal{K}_i} N_k(k_\perp,k_\parallel) \times \Delta^2_{21} (k_\perp, k_\parallel), 
    \end{align}
    where $\mathcal{K}_i$ is the set of $k = \sqrt{k_\perp^2 + k_\parallel^2} \in k_i$\footnote{For more details, see \texttt{cylindrical\_to\_spherical} in \hyperlink{https://github.com/21cmfast/tuesday}{tuesday}.}. 
\label{lastpage}
\end{document}